\documentclass{aa}
\usepackage{graphicx}
\usepackage{txfonts}
\usepackage[figuresright]{rotating}
\usepackage{natbib}
\bibpunct{(}{)}{;}{a}{}{,}

\def\ngc{NGC\,6231}
\def\hda{HD\,152248}
\def\cpd{CPD\,$-$41\degr7742}
\def\sco{Sco\,OB\,1}

\def\ic{IC\,4628}
\def\dsco{$\zeta^1$\,Sco}
\def\tr{Tr\,24}
\def\bceph{$\beta$\,Cephei}

\def\halph{H$\alpha$}
\def\hbet{H$\beta$}

\def\cnts{cnt\,s$^{-1}$}
\def\ergs{erg\,s$^{-1}$}
\def\ergscm{erg\,cm$^{-2}$\,s$^{-1}$}

\def\msol{M$_{\odot}$}

\def\chandra{\emph{Chandra}}
\def\rosat{\emph{{\sc ROSAT}}}

\def\xmm{{\sc XMM}\emph{-Newton}}

\def\epic{{\sc EPIC}}
\def\mos{{\sc MOS}}
\def\pn{pn}
\def\rgs{{\sc RGS}}
\def\epicmos{{\sc EPIC MOS}}
\def\epicpn{{\sc EPIC} pn}

\def\mekal{{\sc mekal}}

\def\eml{{\it emldetect}}
\def\sbl{SSB06}

\defcitealias{SBL98}{SBL98}

\begin{document}
   \title{An \xmm\ view of the young open cluster NGC~6231\thanks{Based on observations collected with \xmm, an ESA science mission with instruments and contributions directly funded by ESA Member States and the USA (NASA).}}
   \subtitle{I. The catalogue \thanks{Tables \ref{tab: Xcat} and \ref{tab: Xcat_iden} are available via the CDS web site: http://cdsweb.u-strasbg.fr .} }

   \author{
          H. Sana\inst{1}\fnmsep\thanks{FNRS Research Fellow (Belgium)},
          E. Gosset\inst{1}\fnmsep\thanks{FNRS Research Associate (Belgium)},
          G. Rauw\inst{1}\fnmsep$^\dagger$,
          H. Sung\inst{2}
          \and
          J.-M. Vreux\inst{1}
          }

   \offprints{H. Sana}

   \institute{Institut d'Astrophysique et de G\'eophysique, University of Li\`ege, All\'ee du 6 Ao\^ut 17, B\^at. B5c, B-4000 Li\`ege, Belgium\\
              \email{sana@astro.ulg.ac.be, gosset@astro.ulg.ac.be, rauw@astro.ulg.ac.be,\\ vreux@astro.ulg.ac.be}
          \and
              Department of Astronomy and Space Science, Sejong University, Kunja-dong 98, Kwangjin-gu, Seoul 143-747, Korea\\
             \email{sungh@sejong.ac.kr}
   }

   \date{Received October 15, 2004; accepted March 16, 2999}

   \abstract{
This paper is the first of a series dedicated to the X-ray properties of the young open cluster \ngc. Our data set relies on an \xmm\ campaign of a nominal duration of 180\,ks and reveals  that  \ngc\  is very rich in the X-ray domain too. Indeed, 610 X-ray sources are detected in the present field of view, centered on the cluster core. The limiting sensitivity of our survey is  approximately  $6\times10^{-15}$\ergscm\ but clearly depends on the location in the field of view and on the source spectrum. Using different existing catalogues, over 85\% of the X-ray sources could be associated with at least one optical and/or infrared counterpart  within a limited cross-correlation radius of 3\arcsec\ at maximum. The surface density distribution of the X-ray sources presents a slight N-S elongation. Once corrected for the spatial sensitivity variation of the \epic\ instruments, the radial profile of the source surface density is well described by a King profile with a central density of about 8  sources per arcmin$^2$ and a core radius close to 3.1\,arcmin. The  distribution of the X-ray sources seems closely related to the optical source distribution. The expected number of foreground and background sources should represent about 9\% of the detected sources, thus strongly suggesting that most of the observed X-ray emitters are physically belonging to \ngc. Finally, beside a few bright but soft objects -- corresponding to the early-type stars of the cluster -- most of the sources are relatively faint ($\sim5\times10^{-15}$~\ergscm) with an  energy distribution peaked around 1.0-2.0\,keV. 
   \keywords{
     Open clusters and associations: individual: NGC 6231 --
     X-rays: individuals: NGC 6231 --
     X-rays: stars --
     Stars: early-type -- Catalogues}
   }

   \titlerunning{An \xmm\ view of \ngc. I.}
   \authorrunning{H.~Sana et al.~}
   \maketitle
%
%________________________________________________________________

\section{Introduction }\label{sect: intro}

Detailed studies of young clusters are powerful tools to probe crucial astrophysical issues. Because they {\it a priori} contain both early-type stars and pre-main sequence (PMS) stars, young clusters are privileged laboratories to test star formation and evolution theories. They indeed  provide a homogeneous sample of stars in terms of distance, reddening, environment, chemical composition and age. With the currently available X-ray observatories, unprecedented investigations of young open clusters in the X-ray domain have been performed in the past few years. The increased sensitivity, spectral power and resolution of the \xmm\ and \chandra\ observatories, compared to X-ray satellites of the previous generations, give now a much more complete view of the X-ray properties of the star populations in clusters.

For example, a 76\,ks \chandra\ observation of the embedded young cluster NGC\,2024 ($d\sim410$\,pc; age $=$ 0.3 -- a few Myr) revealed 283 X-ray sources displaying  heavily absorbed hard spectra with a mean temperature k$T\sim3$\,keV \citep{SGB03}. A significant fraction (25\%) of the X-ray sources shows a wide range of variability within the exposure duration. In addition, \chandra\ detected at least 96\% of the known classical T Tauri stars in NGC\,2024. Results for other clusters are very similar. \citet{RdBG03} performed a 20\,ks observation of NGC\,6383 ($d\sim1.4$\,kpc; age $=$ 1.7 -- 5\,Myr) and found 77 sources, mostly centered on the cluster location. An important fraction of these sources are probable PMS objects. Using both \xmm\ and \chandra\ facilities, \citet[ and references therein]{PZ04} studied the very young stellar cluster IC\,348 ($d\sim310$\,pc; age $\sim$ 2\,Myr) and found 286 X-ray sources among which over 50 classical T Tauri stars. Comparison of \chandra- and \xmm- based spectral properties suggested that the X-ray characteristics of T Tauri stars remain mostly constant over periods of years. NGC\,6530 ($d\sim1.8$\,kpc; age $\sim$ 1.5--2.0\,Myr) is a very rich open cluster containing several massive O-type stars as well as a large population of B-type stars. \xmm\ observations by \citet{RNG02} revealed 119 sources, of which a large fraction are PMS candidates. Similarly to \citeauthor{SGB03} results, the X-ray spectra of the PMS candidates are characterized by temperatures of a few keV. Using a 60\,ks \chandra\, observation centered on the same cluster, \citet{DFM04} revealed 884 X-ray sources, among which 90 to 95\% are PMS stars.  

From this review of the recent literature, there is an obvious body of observations showing that, besides the expected X-ray emission from massive stars, a large population of X-ray emitting low-mass PMS stars is to be found while observing young clusters. The present study of the very rich cluster \ngc\ lies in this framework. It aims at a better comprehension of both early-type stars and young open clusters by extending the previous sample of investigations to clusters with a large O-type star population. A severe limitation of several of the above cited works is the lack of detailed studies on the concerned cluster at other wavelengths. Indeed, with \chandra\ and \xmm, the X-ray observations are so deep that a deep optical photometry of the field is required. Such a data set is indeed essential to, for example, more clearly identify the evolutionary status of the different sub-populations of the cluster. Fortunately, as shown by the literature review of Sect. \ref{sect: lit}, the  stars in \ngc\ have been thoroughly studied. Together with the depth of the present X-ray campaign, this  is one of the strengths of the current work. Finally, the present work distinguishes itself from the previous investigations because of the particular planning of the X-ray observations. Indeed our \xmm\ campaign towards NGC\,6231 was actually split into six successive pointings, spread over a period of five days. This will allow us to probe the variability of the X-ray emission of the detected sources on different time-scales.

A detailed analysis of the central target of the field, the colliding wind binary \hda, has been presented recently in a dedicated paper \citep{SSG04}. The source will therefore not be further discussed in more details in the present paper. Preliminary results from this campaign, mainly related to the early-type X-ray emitters, were also presented in \citet{SRG05, SGR06_elescorial, SNG06}.
In the present paper, we focus on the X-ray catalogue and we discuss some general properties of the detected sources. Other aspects of the X-ray properties of \ngc, such as the early-type and the pre-main sequence population characteristics, will be addressed in subsequent papers of this series.

This first paper is organized as follows. After a review of the abundant literature on \ngc\ and on the  \sco\ association, Sect. \ref{sect: obs} describes the campaign and the subsequent data reduction processes. In Sect. \ref{sect: Xcatal}, we address the detection and identification of the sources in the \xmm\ field of view, and we present the resulting X-ray catalogues. Finally,  we probe the main properties of the cluster X-ray emitters (Sect. \ref{sect: Xsources}). Sect. \ref{sect: ccl} summarizes the results of the present work.

\section[]{\ngc\ and the \sco\ association: a literature review }\label{sect: lit}

\subsection{The \sco\ association}

Located in the Sagittarius-Carina spiral arm of our galaxy ($\alpha(2000)=16^\mathrm{h}53\fm6$, $\delta(2000)=-41\degr57'$; $l=343\fdg3$, $b=1\fdg2$, \citealt{PHC91}), the  \sco\ association is an extremely rich and interesting region of the sky. 2\degr\ long by 1\degr\ wide, it extends from the gaseous nebula \ic\ on its northern end to the young open cluster \ngc\ towards its southern end. Its major axis is approximately parallel to the Galactic plane \citep{MGG53}. A sparser group, \tr, is to be found near \ic\ while two other clusters, NGC\,6242 and NGC\,6268, lie slightly north of the association. Finally the  H\,{\sc ii}\ region IC\,4878, centered on \ngc, extends by about 4\degr\ by 5\degr\ in the form of an elliptical ring and is probably triggered by the cluster. The emission nebula is faint within the ring but is very bright where the ring is crossed by the northern end of the association \citep{BBG66}. \\ 

 The interest in \sco\ mainly originates from its extended early-type star content (\citealt{Sh83}; \citealt{RCB97}). Beyond the numerous O- and B-type stars, the association also shelters two of the rare Wolf-Rayet (WR) stars, two Of stars displaying P Cygni profiles as well as several \bceph\ variables \citep{BaE85, ASK01}. Among the peculiar objects found within \sco\ is the bright star \dsco. With an absolute magnitude around $M_\mathrm{V}$=$-$8.3, \dsco\ is one of the brightest stars of the Milky Way. Many of the `normal' early-type stars further present signs of variability and have a good chance to be binary  systems \citep[e.g.][]{Rab96,ASK01}.\\

\subsection{The \ngc\ cluster}

Located near the southern end of the \sco\ association, the young  open cluster \ngc\ ($\alpha(2000)=16^\mathrm{h} 54^\mathrm{m} 09^\mathrm{s}$, $\delta(2000)=-41\degr49'36''$) contains an important number of bright early-type stars in its centre. Often considered as the nucleus of the association \citep{BBG66}, its relationship to \sco\ has been subject to different interpretations with time. 
Though \citet{HW84} presented evidence that \tr\ and \sco\ form a single aggregate, these authors proposed that \ngc\ is actually a foreground cluster. \citeauthor{HW84} also found a sub-cluster of PMS stars in the vicinity of \tr. 
 Based on an extensive set of data, \citet{PHYB90,PHC91} re-addressed these issues and carefully studied the interrelation between the three aggregates. They established that \sco, \ngc\ and \tr\ are located at the same distance and have the same age, thus demonstrating that \ngc\ is not a foreground object but is clearly embedded in the \sco\ association. \ngc\ therefore retains its status as the nucleus of the association. \citeauthor{PHC91} could however not confirm the three stellar sub-aggregates found by \citet{Se68b} in \tr\ and, as suggested by \citet{HW84}, they casted further doubts on the physical reality of the \tr\ aggregate itself. \citeauthor{PHC91} finally confirmed the existence of a PMS sub-cluster near \tr. \\

\begin{table*}
\centering
\caption{Journal of the \xmm\ observations of \ngc. Columns 2 and 3 give the spacecraft revolution number and the observation ID. The Julian Date (JD) at mid-exposure is reported in Col. 4. Cols. 5 to 7 (resp. Cols. 8 to 10)  list the performed (resp. effective -- i.e. corrected for the high background periods) exposure times for the  \epicmos1, \epicmos2 and \epicpn\  instruments. The last column provides the position angle (PA) associated to the revolution. The total campaign duration is given in the last line of the table.
}
\label{tab: journal} 
\centering          

\begin{tabular}{c c c c c c c c c c c c}
\hline
\hline
Obs. \# & Sat. & Exposure &     JD & \multicolumn{3}{c}{Performed duration (ks)}& \multicolumn{3}{c}{Effective duration (ks)} & PA \\
       & Rev. &  ID      &JD$-2\,450\,000$& \mos1 & \mos2 & \pn  & \mos1 & \mos2 & \pn  & DDD:AM:AS.s \\
\hline
1 & 319 & 0109490101 & 2158.214 & 33.3 & 33.3 & 30.7 & 33.1 & 33.2 & 30.6 & 274:57:11.5 \\
2 & 319 & 0109490201 & 2158.931 & 22.1 & 22.1 & 20.2 & 19.8 & 19.8 & 16.5 & 274:57:11.5 \\
3 & 320 & 0109490301 & 2159.796 & 34.4 & 34.4 & 31.8 & 33.7 & 33.9 & 30.1 & 275:35:26.6 \\
4 & 320 & 0109490401 & 2160.925 & 31.4 & 31.4 & 29.1 & 26.0 & 24.3 & 11.7 & 275:35:26.6 \\
5 & 321 & 0109490501 & 2161.774 & 31.1 & 31.1 & 28.5 & 30.9 & 31.0 & 28.4 & 276:13:34.9 \\
6 & 321 & 0109490601 & 2162.726 & 32.9 & 32.9 & 30.3 & 32.9 & 32.8 & 30.3 & 276:13:34.9 \\
\hline
\multicolumn{3}{l}{Total duration (ks)} & & 185.2 & 185.2 & 170.6 & 176.5 & 175.0 & 147.5 & \\
\hline
\end{tabular}
\end{table*}

The properties of \ngc\ and of its stellar content have been thoroughly investigated during the past century. Three main streams of investigation were indeed designed, namely photometry, spectral classification and radial velocity measurements. The photometric approach is however predominant and was extensively performed using different photometric systems. The bulk of the available literature on the cluster relies on photographic,  photoelectric or CCD campaigns: \citet[ PV]{BC53}, \citet{Ho56}, \citet[ Walraven]{WW60}, \citet[ UBV]{FSTW61}, \citet[ PV]{BK63}, \citet[ UBV\,\hbet]{BBG66}, \citet[ UBV]{FF68}, \citet[ UBV]{Se68a}, \citet[ UBV]{SHS69}, \citet[ uvby\,\hbet]{CBHP71},  \citet[ UBV]{GaS79}, \citet[ uvby\,\hbet]{Sh83}, \citet[ UBV]{HW84}, \citet[ Walraven]{vGBvG84}, \citet[ uvby]{PHC91}, \citet[ UBV]{MMM93}, \citet[ uvby\,\hbet]{BL95}, \citet[ Geneva]{RCB97}, \citet[ UBV(RI)$_\mathrm{C}$\,\halph]{SBL98}, \citet[ UBVI$_\mathrm{C}$]{BVF99}. The more recent works (from $\sim$1990's) offer a much more complete view of the cluster both in terms of their angular extent and of the magnitude limit reached. An  extensive still careful comparison of most (if not all) the works published prior to 1990 has been performed by \citet{PHC91}. \\

Spectral classification of the cluster objects has mainly been carried out by \citet{MWC53}, \citet{Ho56}, \citet{FSTW61}, \citet{SHS69},  \citet{GaS79},  \citet{LM80},  \citet{CA71},  \citet{Wal72},  \citet{Mat88,Mat89},  \citet{GM01} and \citet{phd}. Radial velocity campaigns were led essentially by \citet{Str44}, \citet{HCB74}, \citet{LM83}, \citet{LMGM88}, \citet{PHYB90}, \citet{PBG94}, \citet[ IUE data]{SL01}, \citet{GM01}, \citet{SRG02} and \citet{phd}. Several authors also paid a special attention to particular objects, mainly binaries of which they performed a more detailed study. These objects are WR\,79 \citep{Lu97}, HD\,152218 \citep{SLP97}, HD\,152248 (\citealt{SLP96}; \citealt{PGB99}; \citealt{SRG01}; \citealt{SSG04}),  \cpd\ \citep{SHRG03, SAR05} and HD\,152219 \citep{SGR06_219}. \\

Aside from these three main streams, several authors addressed specific aspects of the cluster that provide a useful complementary view. Among other topics, photometric variability of a few dozens of  objects was investigated by  \citet{Bal83}, \citet{BaE85}, \citet{Bal92} and more recently by \citet{ASK01}. These studies allowed to detect several \bceph, a couple of $\delta$ Scuti and a few other variable stars, including a couple of eclipsing binaries. The binary fraction was estimated by \citet{Rab96} and \citet{GM01}. \citeauthor{Rab96} derived a minimum binary frequency of 52\% in his sample of 53 B-type stars with a spectral type between B1 and B9 while \citet{GM01} obtained an extremely high frequency of 82\% for stars earlier than B1.5V and, in particular, of 79\% for the O-type stars of the cluster.  \citet{RM98} showed evidence of mass segregation in \ngc, most probably related to the formation processes rather than to the dynamical  evolution of the cluster.  Proper motions were studied by \citet{Br67} and \citet{Lav72} while most of the O-type stars of the clusters were included in the large ICCD Speckle campaign of \citet{MGH98}. \\

The distance modulus ($DM$) of the cluster reported in the earlier literature ranges from 10.7 \citep{Me81} to 11.9 \citep[ 2300pc -- cited by \citealt{BBG66}]{Ho56}. In a more recent work, \citet{PHC91} obtained $DM=11.50$ and 11.55 for \sco\ and \ngc\ respectively, with an uncertainty of about 0.32. \citet{BL95} derived $DM=11.08 \pm 0.05$ for \ngc; \citet{RCB97}, $11.2\pm0.4$; \citet{SBL98}, $11.0\pm0.07$ and \citet{BVF99} $11.5\pm0.25$. The weighted mean of these five latter values gives $DM=11.07 \pm 0.04$, corresponding to an actual distance of $1637 \pm 30$\,pc. The same authors (but \citeauthor{SBL98}) respectively derived ages of $7.9\pm0.9$\,Myr, $5\pm1$\,Myr, $3.8\pm0.6$\,Myr and 3 to 5\,Myr. On the basis of the R-H$\alpha$ index, \citet{SBL98} found 12 PMS stars plus 7 PMS candidates. \\

A controversial question is the probable differential reddening across the cluster. Such a differential reddening was first suggested by \citet{BK63}, outlining that the southern part of the cluster suffers a heavier reddening. Other authors \citep{Sh83, PHC91, BL95} rather proposed a uniform reddening across the field. More recently, \citet{RCB97} and \citet{SBL98} results strongly support the first idea of \citet{BK63}, and \citeauthor{SBL98}  presented a  map of the reddening distribution in \ngc. There seems to be an agreement in the early literature that most of the reddening occurs between a distance of 100 and 1300~pc.  Based on FUSE observations, \citet{MBdB04} recently confirmed angular variations in the column density towards the core of the cluster. They reported that the absorption towards \ngc\ occurs in several foreground clouds. The main absorption component lies in the Lupus cloud region at a distance of 150~pc, while the second one is probably in the vicinity of the \sco\ shell surrounding \ngc. Finally, \citet{Cr01} probed the structure of the interstellar Na\,{\sc i} and K\,{\sc i} towards the cluster and reached conclusions similar to those of \citet{MBdB04}. \citeauthor{Cr01} also outlined that no clues of active shocks in the shell components could be found. Polarimetric observations were performed by \citet{FMV03} who found evidence for a past supernova explosion in the cluster. These authors however suggested that their observations could also be explained by a bubble triggered by  winds from hot stars. \\

   \begin{figure*}
   \centering
   \includegraphics[width=\textwidth]{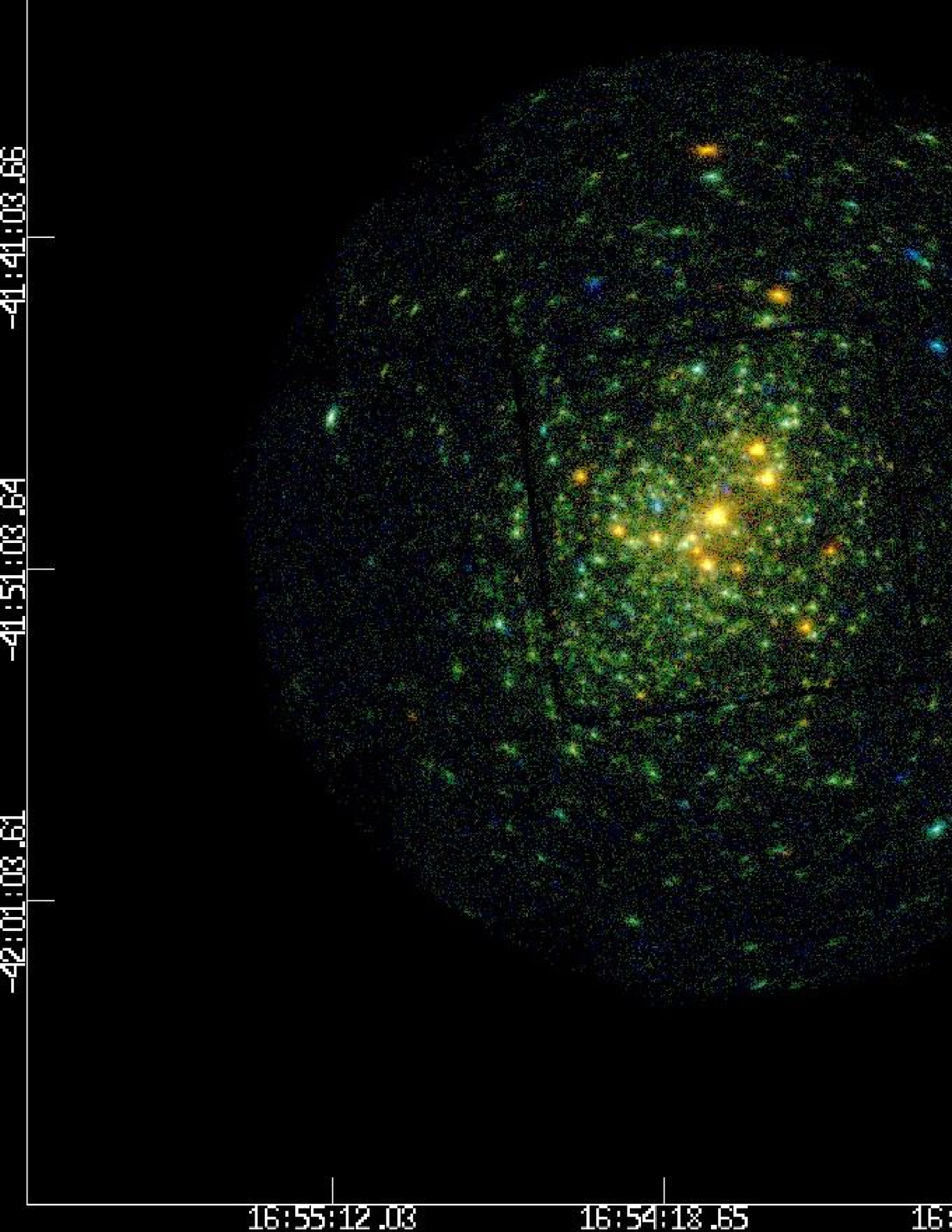}
   \caption{Combined \epicmos\ three-color X-ray image of the young open cluster \ngc. The field is roughly 30\arcmin\ in diameter and is centered on \hda. North is up and East is to the left. The different colors correspond to different energy ranges: red: 0.5-1.0\,keV; green : 1.0-2.5\,keV; blue : 2.5-10.0\,keV.              }
         \label{fig: ngc6231}
   \end{figure*}

Finally, we note that the Hipparcos parallaxes derived for \ngc\ were known to be problematic with a negative mean value of $-0.8\pm0.4$\,mas \citep{AL99}. These results were recently revised by \citet{Ma03} who obtained $1.7 \pm0.4$\,mas, corresponding to a distance modulus of $8.9\pm0.5$, however still far from the mean value obtained from the photometric studies.

Turning to the X-ray domain, \ngc\ was observed by the ROentgen SATellite (\rosat). Thirty-five objects were detected, mainly associated with the early-type stars of the cluster.  \citet{Cor96, Cor99} presented some results of this campaign as well as the X-ray light curve of three objects, namely HD\,152218, \hda\ and HD\,152249. Only \hda\  displayed clear variations of its flux.   Finally a few objects were also observed at radio wavelength \citep{GCS02,GCS03} but only half of them were detected.\\

%__________________________________________________________________

\section[]{Observations and Data Reduction }\label{sect: obs}

\subsection{The {\rm XMM}-Newton campaign \label{ssect: obs}}

The \xmm\ campaign towards \ngc\ has already been described in \citet{SSG04}. For the sake of completeness, we again give here a brief description of the X-ray observations. In September 2001, during satellite revolutions 319 to 321, the \xmm\ observatory \citep{Jansen01_xmm} performed six successive exposures of an approximate  duration of 30\,ks. The field of view (FOV) was centered on the O7.5\,III+O7\,III colliding wind binary \hda\ \citep[$\alpha_{2000}=16^\mathrm{h} 54^\mathrm{m} 10$\fs06, $\delta_{2000}=-41$\degr49\arcmin 30\farcs 1;][]{SRG01}, in the core of the cluster. Position angles (PAs) were very similar through the six exposures, ranging approximatively from 274\fdg95 to 276\fdg23. All three EPIC instruments \citep{Struder01_pn_mnras,Turner01_mos_mnras} were operated in the Full Frame mode together with the Thick Filter to reject UV/optical light. The \rgs\ spectrographs \citep{denHerder01_rgs_mnras} were run in the Standard Spectroscopic mode. Due to the brightness of the cluster objects in the FOV, the Optical Monitor \citep{Mason01_om_mnras} was switched off throughout the campaign. Table \ref{tab: journal} provides the journal of the X-ray observations.

\subsection{Data Reduction }\label{ssect: reduc}

The \epic\ Observation Data Files (ODFs) were processed using the XMM-Science Analysis System (SAS) v\,5.4.1 implemented on our computers in Li\`ege. We applied the {\it emproc} and {\it epproc} pipeline chains respectively to the \mos\ and \pn\ raw data to generate proper event list files. No indication of pile-up was found in the data. We then only considered  events with patterns 0-12 (resp. 0-4) for \mos\ (resp. \pn) instruments and we applied the filtering criterion XMMEA\_EM (resp. FLAG\,=\,0) as re\-commended by the Science Operation Centre (SOC) technical note XMM-PS-TN-43\,v3.0. For each pointing, we rejected periods affected by soft proton flares. For this purpose, we built light curves at energies above 10\,keV\footnote{Expressed in Pulse Invariant (PI) channel numbers and considering that 1 PI channel approximately corresponds to 1 eV, the adopted criterion is actually PI $>$ 10\,000.} and discarded high background observing periods on the basis of an empirically derived threshold (adopted as equal to 0.2 and 1.0~\cnts\ for the \mos\ and \pn\ instruments respectively). The so-defined GTIs (Good Time Intervals) were used to produce  adequate X-ray event lists for each pointing from which we extracted images using x- and y-image bin sizes of 50 virtual pixels \footnote{Though the physical pixels of the \epicmos\ and \pn\ detectors have an extent on the sky of respectively 1\farcs1 and 4\farcs1, the virtual  pixels of the three instruments correspond to an extent 0\farcs05. The obtained images have thus a pixel size of 2\farcs5.}.

We finally combined the event lists obtained for all six pointings to increase the statistics of faint sources. For this purpose, we used the SAS task {\it merge}. For each \epic\ instrument, we included the event lists resulting from different pointings one by one. We also built merged event lists that combine the twelve \mos\  or the eighteen \epic\ event lists. The Attitude Files generated by the pipeline were merged using the same approach and we adopted, for handling the merged event lists, the Calibration Index File (CIF) and the ODF corresponding to the first pointing (Obs. 1 in Table \ref{tab: journal}). \\

%__________________________________________________________________

\section[]{X-ray source detection and identification} \label{sect: Xcatal}
In this section, we focus  on the  detection and identification of the X-ray sources in the \xmm\ FOV. For this purpose, we only used the merged event lists and images, accounting in this way for the six pointings at once. The total effective exposure times towards the cluster are, respectively for the \mos1, \mos2\ and \pn\ instruments, of 176.5, 175.0 and 147.5\,ks. Together with the high sensitivity of the \xmm\ observatory, the combination of the six pointings and of the three instruments provides one of the deepest X-ray views of a young open cluster. Fig.~\ref{fig: ngc6231} shows a three-colour image of \ngc\ and reveals a densely populated field with hundreds of point-like X-ray sources. This section therefore aims at providing a uniform catalogue of these sources. It is organised as follows. First we present the source detection procedure as well as a brief description of the obtained catalogues. As a next step, we focus on the identification of the X-ray sources and, finally, we investigate the detection limit of the present data set.

\subsection{Source Detection }\label{ssect: source_detect}

We based our source detection on the SAS detection chain {\it edetect\_chain}. For this purpose, we selected three energy ranges, a soft (S$_\mathrm{X}$) band (0.5-1.0\,keV), a medium (M$_\mathrm{X}$) band (1.0-2.5\,keV) and a hard (H$_\mathrm{X}$) band (2.5-10.0\,keV), and we built the corresponding input images for the different instruments. The {\it edetect\_chain} task is formed by the succession of the SAS tasks {\it eexpmap}, {\it emask}, {\it eboxdetect} run in local mode, {\it esplinemap}, again {\it eboxdetect} run in map mode and finally {\it emldetect}:
\begin{enumerate}
\item[-] {\it eexpmap} calculates the exposure maps corresponding to the input images;
\item[-] these exposure maps are used by {\it emask} to build masks which select the relevant image areas where the detection should take place;
\item[-] {\it eboxdetect}, run in local mode,  uses a 5$\times$5 pixel box and a surrounding  background area to search for significant sources simultaneously in all input images;
\item[-] {\it esplinemap} uses the resulting source list to remove sources from the input images and creates smooth background maps by fitting a 2-D spline to the source-subtracted images;
\item[-] run in map mode, {\it eboxdetect} uses a 5$\times$5  pixel box and the values from the background maps to search for significant sources simultaneously in all input images;
\item[-] {\it emldetect} finally uses the preliminary source list from {\it eboxdetect} and determines the source parameters (e.g. coordinates, count rates, hardness ratios, etc.) by means of simultaneous maximum likelihood psf (point spread function) fitting to the source count distribution in all energy bands of each EPIC instrument. It also provides an equivalent logarithmic likelihood $L_2$ (Eq.~\ref{eq: L2}) commonly used as an indication of the {\it reality} of the corresponding source.
\end{enumerate}

 From our experience, the {\it eboxdetect} task run in map mode  tended to eliminate some apparently real sources from the intermediate source list. We therefore preferred to use the preliminary source list obtained by {\it eboxdetect} in local mode as an input list for the psf fitting step performed by the {\it emldetect} task. This approach does not bias the result since, if the source is real, the psf fitting will provide a large logarithmic likelihood while, if instead the source is fake, the logarithmic likelihood will be low and the source will be rejected. Though more expensive in computation time, this approach results in a more complete source list. As it was known that the equivalent logarithmic likelihood values ($L_2$) computed by the {\it emldetect} task in the SAS\,v\,5.4.1 (and earlier versions) were erroneous, we implemented a patch to recover the correct $L_2$ values. We give a brief description of it in Appendix \ref{app: L2_a}. The problem is now fixed from SAS version 6.0 on. We checked our corrected logarithmic likelihood values against SAS\,v\,6.0 results and found them to be in close agreement.\\

 We first performed single psf fit detection but, due to the crowdedness of the field, we also allowed for simultaneous fitting of up to four sources. In doing so, we adopted a value of 0.68 for the two parameters {\it scut} and {\it ecut}. This choice was led by the need to account for as large an energy fraction of the psf as possible while keeping the computation time down to reasonable limits. Due to the densely populated field, the wings of the source psf are often largely contaminated by emission from neighbouring sources. The adopted values therefore appeared as a reasonable compromise. On the axis, this corresponds to a physical radius of about 15\arcsec. Only a few tens of sources actually required multi-psf fitting, with three psf being simultaneously adjusted at the maximum. Finally, we re-ran the {\it emldetect} task allowing for extended sources to be fitted. A careful comparison of the resulting lists shows that only a few sources increase significantly their detection likelihood while allowing for extended source fitting. An inspection of the X-ray images and of the optical catalogues reveals that these sources most probably correspond to unresolved point-like sources rather than to physically extended sources. \\

\begin{table}
\caption{Adopted detection thresholds for the equivalent logarithmic likelihood $L_2$ corresponding to the different \epic\ instruments (left column) or to any combination of them (right column). Appendix \ref{app: L2_b} provides more details on how these values were computed.}
\label{tab: l2}
\centering
\begin{tabular}{c c | c c}
\hline
\hline
\vspace*{-3mm} & & &  \\
Instr. & $L_2$ & Instr. Comb. & $L_2$ \\
\vspace*{-3mm} & & &  \\
\hline
\mos1 & 11 & \mos1 + \mos2  & 21 \\
\mos2 & 11 & \mos1 + \pn    & 35 \\
\pn   & 25 & \mos2 + \pn    & 35 \\
      &    & \mos1 + \mos2 + \pn & 45 \\
\hline
\end{tabular}
\end{table}

%%%%%%%%%%%%%%%%%%%%%%%%%%%%%%%%%%%%%%%%%%%%%%%%%%%%%%%%%%%%%%%%%%%%%%%%%%%%%%%
%%%%%%%%%%%%%%%%%%%%%%%%% POINT-LIKE SOURCE CATALOGUE %%%%%%%%%%%%%%%%%%%%%%%%%
%%%%%%%%%%%%%%%%%%%%%%%%%%%%%%%%%%%%%%%%%%%%%%%%%%%%%%%%%%%%%%%%%%%%%%%%%%%%%%%

\begin{sidewaystable*}
\begin{minipage}[t][180mm]{\textwidth}
\caption{Sample of the \ngc\ X-ray source catalogue. Each part of the table refers to a particular \epic\ instrument. First column gives the source number. Second column provides the source name (based on position in J2000.0) following the format specified for \xmm\ sources. Note that the coordinates are truncated, not rounded. The estimated error on the 2D position (in \arcsec) is listed in the third column ($\sigma_{\alpha\delta}$). The first three columns are reproduced at the beginning of the three parts of this table. Each sub-table concerns one of the three \epic\ instruments and follows the same arrangement. Cols. 4-16 (resp. 17-29 and 30-42) give the equivalent logarithmic likelihood $L_2$ for the given instrument, the total count rate $cr$ in the whole energy band (0.5-10.0\,keV) and its associated error ($\sigma$), the count rates in the different energy bands (S$_\mathrm{X}$: [0.5-1.0\,keV], M$_\mathrm{X}$: [1.0-2.5\,keV], H$_\mathrm{X}$: [2.5-10.0\,keV]) and their errors and, finally, the two hardness ratios $HR_1$ and $HR_2$ (Eqs. \ref{eq: hr1} and \ref{eq: hr2}) as well as their related errors ($\sigma_{HR1}$ and $\sigma_{HR2}$). The count rates and the related uncertainties are all expressed in $10^{-3}$~\cnts. Col. 43 provides the total equivalent logarithmic likelihood $L_2^\mathrm{EPIC}$ that has been compared to the adopted detection limit (see Table \ref{tab: l2}) according to the combination of instruments used (Col. 44). For those sources for which the fit is improved while adjusting an extended psf model, a note ('ext') is stated in Col. 45. Finally Col. 46 provides a warning for some sources for which no instrument combination were found to be far enough from gaps, instrument edges or detector bad columns. The electronic version of the table further provides two additional columns that give the derived source coordinates with a precision of 0\fs01 in right ascension and of 0\farcs1 in declination. The full table can be obtained from the Centre de Donn\'ees astronomiques de Strasbourg (CDS, http://cdsweb.u-strasbg.fr). }
\label{tab: Xcat}
\tiny
\begin{tabular}{c c c | c c c c c c c c c c c c c | c c c c}
\hline 
\hline
  \multicolumn{3}{c|}{XMMU J} &\multicolumn{13}{c|}{\epicpn\ instrument}  \\
X\# & HHMMSS.s$\pm$DDAMAS & $\sigma_{\alpha\delta}$ & 
$L_2^\mathrm{pn}$ & cr$_\mathrm{pn}$ & $\sigma_\mathrm{pn}$ & 
cr$^\mathrm{S}_\mathrm{pn}$ & $\sigma^\mathrm{S}_\mathrm{pn}$ & 
cr$^\mathrm{M}_\mathrm{pn}$ & $\sigma^\mathrm{M}_\mathrm{pn}$ & 
cr$^\mathrm{H}_\mathrm{pn}$ & $\sigma^\mathrm{H}_\mathrm{pn}$ & 
$HR_1^\mathrm{pn}$  & $\sigma_{HR1}^\mathrm{pn}$ &
$HR_2^\mathrm{pn}$  & $\sigma_{HR2}^\mathrm{pn}$ \\ 

[1] & [2] & [3] & [4] & [5] & [6] & [7] & [8] & [9] & [10] & [11] & [12] & [13] & [14] & [15] & [16]  \\
\hline
  1 & 165300.0$-$415444 &   1.2  &       --- &        --- &        --- &        --- &        --- &        --- &        --- &        --- &        --- &        --- &        --- &        --- &        --- &\\ 
  2 & 165304.3$-$415334 &   1.4  &       --- &        --- &        --- &        --- &        --- &        --- &        --- &        --- &        --- &        --- &        --- &        --- &        --- &\\ 
  3 & 165305.2$-$415204 &   1.1  &       --- &        --- &        --- &        --- &        --- &        --- &        --- &        --- &        --- &        --- &        --- &        --- &        --- &\\ 
  4 & 165306.9$-$414930 &   1.2  &       --- &        --- &        --- &        --- &        --- &        --- &        --- &        --- &        --- &        --- &        --- &        --- &        --- &\\ 
  5 & 165307.4$-$414659 &   1.0  &       --- &        --- &        --- &        --- &        --- &        --- &        --- &        --- &        --- &        --- &        --- &        --- &        --- &\\ 
  6 & 165307.4$-$414345 &   0.5  &       --- &        --- &        --- &        --- &        --- &        --- &        --- &        --- &        --- &        --- &        --- &        --- &        --- &\\ 
  7 & 165308.1$-$414533 &   0.9  &       --- &        --- &        --- &        --- &        --- &        --- &        --- &        --- &        --- &        --- &        --- &        --- &        --- &\\ 
  8 & 165310.2$-$414733 &   1.1  &       --- &        --- &        --- &        --- &        --- &        --- &        --- &        --- &        --- &        --- &        --- &        --- &        --- &\\ 
  9 & 165310.7$-$414451 &   1.0  &       --- &        --- &        --- &        --- &        --- &        --- &        --- &        --- &        --- &        --- &        --- &        --- &        --- &\\ 
 10 & 165311.6$-$414755 &   0.6  &      348.9 &      7.427 &      0.492 &      2.271 &      0.247 &      3.871 &      0.301 &      1.285 &      0.301 &      0.261 &      0.062 &     $-$0.501 &      0.092 &\\ 
 11 & 165313.0$-$415049 &   0.5  &      146.0 &      4.157 &      0.404 &      1.214 &      0.195 &      2.367 &      0.257 &      0.577 &      0.243 &      0.322 &      0.087 &     $-$0.608 &      0.137 &\\ 
 12 & 165313.2$-$415222 &   0.4  &      295.3 &      6.215 &      0.438 &      1.623 &      0.211 &      3.538 &      0.279 &      1.054 &      0.264 &      0.371 &      0.065 &     $-$0.541 &      0.093 &\\ 
 13 & 165313.5$-$415133 &   0.6  &      241.9 &      5.432 &      0.431 &      2.037 &      0.218 &      2.718 &      0.261 &      0.677 &      0.265 &      0.143 &      0.070 &     $-$0.601 &      0.129 &\\ 
 14 & 165315.3$-$415011 &   0.8  &       65.5 &      2.567 &      0.344 &      0.787 &      0.157 &      1.319 &      0.198 &      0.462 &      0.234 &      0.253 &      0.117 &     $-$0.481 &      0.203 &\\ 
\hline
\vspace*{0mm}\\
\hline
\hline
   \multicolumn{3}{c|}{XMMU J} & \multicolumn{13}{c|}{\epicmos1 instrument} \\
X\# & HHMMSS.s$\pm$DDAMAS & $\sigma_{\alpha\delta}$ & 
$L_2^\mathrm{MOS1}$ & cr$_\mathrm{MOS1}$ & $\sigma_\mathrm{MOS1}$ & 
cr$^\mathrm{S}_\mathrm{MOS1}$ & $\sigma^\mathrm{S}_\mathrm{MOS1}$ & 
cr$^\mathrm{M}_\mathrm{MOS1}$ & $\sigma^\mathrm{M}_\mathrm{MOS1}$ & 
cr$^\mathrm{H}_\mathrm{MOS1}$ & $\sigma^\mathrm{H}_\mathrm{MOS1}$ & 
$HR_1^\mathrm{MOS1}$  & $\sigma_{HR1}^\mathrm{MOS1}$ &
$HR_2^\mathrm{MOS1}$  & $\sigma_{HR2}^\mathrm{MOS1}$ \\

[1] & [2] & [3] & [17] & [18] & [19] & [20] & [21] & [22] & [23] & [24] & [25] & [26] & [27] & [28] & [29]  \\
\hline
  1 & 165300.0$-$415444 &   1.2  &       33.0 &      1.079 &      0.199 &      0.318 &      0.081 &      0.636 &      0.120 &      0.125 &      0.136 &      0.334 &      0.140 &     $-$0.672 &      0.303  \\
  2 & 165304.3$-$415334 &   1.4  &       32.0 &      1.137 &      0.192 &      0.241 &      0.083 &      0.591 &      0.112 &      0.305 &      0.132 &      0.420 &      0.161 &     $-$0.319 &      0.213  \\
  3 & 165305.2$-$415204 &   1.1  &       18.8 &      0.911 &      0.191 &      0.189 &      0.073 &      0.400 &      0.110 &      0.322 &      0.138 &      0.357 &      0.206 &     $-$0.108 &      0.252  \\
  4 & 165306.9$-$414930 &   1.2  &        2.0 &      0.338 &      0.148 &      0.001 &      0.027 &      0.000 &      0.048 &      0.338 &      0.138 &     $-$1.000 &    120.824 &      1.000 &      0.283  \\
  5 & 165307.4$-$414659 &   1.0  &       28.9 &      1.048 &      0.178 &      0.000 &      0.035 &      0.287 &      0.092 &      0.761 &      0.148 &      1.000 &      0.246 &      0.453 &      0.149  \\
  6 & 165307.4$-$414345 &   0.5  &      172.8 &      1.957 &      0.192 &      1.627 &      0.159 &      0.330 &      0.099 &      0.000 &      0.043 &     $-$0.663 &      0.089 &     $-$1.000 &      0.259  \\
  7 & 165308.1$-$414533 &   0.9  &       28.2 &      0.921 &      0.179 &      0.199 &      0.063 &      0.566 &      0.119 &      0.157 &      0.118 &      0.481 &      0.146 &     $-$0.566 &      0.266  \\
  8 & 165310.2$-$414733 &   1.1  &       26.8 &      0.598 &      0.137 &      0.393 &      0.082 &      0.206 &      0.088 &      0.000 &      0.065 &     $-$0.312 &      0.215 &     $-$1.000 &      0.629  \\
  9 & 165310.7$-$414451 &   1.0  &       26.4 &      0.726 &      0.147 &      0.021 &      0.034 &      0.623 &      0.108 &      0.083 &      0.094 &      0.935 &      0.103 &     $-$0.766 &      0.237  \\
 10 & 165311.6$-$414755 &   0.6  &       --- &        --- &        --- &        --- &        --- &        --- &        --- &        --- &        --- &        --- &        --- &        --- &        ---  \\
 11 & 165313.0$-$415049 &   0.5  &      116.4 &      1.735 &      0.187 &      0.419 &      0.079 &      1.219 &      0.147 &      0.097 &      0.085 &      0.488 &      0.085 &     $-$0.853 &      0.121  \\
 12 & 165313.2$-$415222 &   0.4  &      112.3 &      1.839 &      0.195 &      0.299 &      0.081 &      1.288 &      0.147 &      0.252 &      0.100 &      0.624 &      0.089 &     $-$0.672 &      0.113  \\
 13 & 165313.5$-$415133 &   0.6  &       71.7 &      1.465 &      0.185 &      0.294 &      0.073 &      0.898 &      0.132 &      0.273 &      0.107 &      0.506 &      0.107 &     $-$0.534 &      0.150  \\
 14 & 165315.3$-$415011 &   0.8  &        8.8 &      0.460 &      0.132 &      0.016 &      0.032 &      0.331 &      0.093 &      0.113 &      0.088 &      0.906 &      0.180 &     $-$0.489 &      0.314  \\
\hline
\vspace*{0mm}\\
\hline
\hline
   \multicolumn{3}{c|}{XMMU J} & \multicolumn{13}{c|}{\epicmos2 instrument} \\
X\# & HHMMSS.s$\pm$DDAMAS & $\sigma_{\alpha\delta}$ & 
$L_2^\mathrm{MOS2}$ & cr$_\mathrm{MOS2}$ & $\sigma_\mathrm{MOS2}$ & 
cr$^\mathrm{S}_\mathrm{MOS2}$ & $\sigma^\mathrm{S}_\mathrm{MOS2}$ & 
cr$^\mathrm{M}_\mathrm{MOS2}$ & $\sigma^\mathrm{M}_\mathrm{MOS2}$ & 
cr$^\mathrm{H}_\mathrm{MOS2}$ & $\sigma^\mathrm{H}_\mathrm{MOS2}$ & 
$HR_1^\mathrm{MOS2}$  & $\sigma_{HR1}^\mathrm{MOS2}$ &
$HR_2^\mathrm{MOS2}$  & $\sigma_{HR2}^\mathrm{MOS2}$ &
$L_2^\mathrm{EPIC}$ & Instr. & Ext. & Comment \\ 

[1] & [2] & [3] & [30] & [31] & [32] & [33] & [34] & [35] & [36] & [37] & [38] & [39] & [40] & [41] & [42] & [43] & [44] & [45] & [46]  \\
\hline
  1 & 165300.0$-$415444 &   1.2  &       --- &        --- &        --- &        --- &        --- &        --- &        --- &        --- &        --- &        --- &        --- &        --- &        --- &       33.0          & m1   \\
  2 & 165304.3$-$415334 &   1.4  &       --- &        --- &        --- &        --- &        --- &        --- &        --- &        --- &        --- &        --- &        --- &        --- &        --- &       32.0          & m1   \\
  3 & 165305.2$-$415204 &   1.1  &       28.0 &      0.833 &      0.152 &      0.183 &      0.074 &      0.599 &      0.110 &      0.051 &      0.074 &      0.531 &      0.159 &     $-$0.844 &      0.212 &       45.9          & mos  \\
  4 & 165306.9$-$414930 &   1.2  &      127.2 &      1.400 &      0.194 &      1.156 &      0.124 &      0.072 &      0.077 &      0.172 &      0.128 &     $-$0.882 &      0.118 &      0.408 &      0.541 &      125.5          & mos  \\
  5 & 165307.4$-$414659 &   1.0  &       16.3 &      0.810 &      0.164 &      0.000 &      0.027 &      0.194 &      0.084 &      0.616 &      0.138 &      1.000 &      0.282 &      0.521 &      0.178 &       44.3          & mos  \\
  6 & 165307.4$-$414345 &   0.5  &      150.1 &      1.869 &      0.202 &      1.409 &      0.146 &      0.460 &      0.121 &      0.000 &      0.067 &     $-$0.508 &      0.105 &     $-$1.000 &      0.293 &      322.1          & mos  \\
  7 & 165308.1$-$414533 &   0.9  &       54.9 &      1.316 &      0.193 &      0.184 &      0.066 &      0.853 &      0.125 &      0.279 &      0.131 &      0.645 &      0.113 &     $-$0.507 &      0.183 &       82.2          & mos  \\
  8 & 165310.2$-$414733 &   1.1  &       14.3 &      0.586 &      0.143 &      0.241 &      0.064 &      0.213 &      0.082 &      0.131 &      0.098 &     $-$0.060 &      0.234 &     $-$0.238 &      0.397 &       40.2          & mos  \\
  9 & 165310.7$-$414451 &   1.0  &       34.5 &      0.989 &      0.173 &      0.296 &      0.075 &      0.567 &      0.117 &      0.126 &      0.103 &      0.315 &      0.148 &     $-$0.635 &      0.251 &       60.1          & mos  \\
 10 & 165311.6$-$414755 &   0.6  &      100.7 &      1.731 &      0.193 &      0.468 &      0.095 &      0.941 &      0.126 &      0.322 &      0.111 &      0.336 &      0.108 &     $-$0.491 &      0.141 &      448.3          & m2pn \\
 11 & 165313.0$-$415049 &   0.5  &       83.1 &      1.280 &      0.160 &      0.263 &      0.068 &      0.983 &      0.130 &      0.034 &      0.065 &      0.578 &      0.097 &     $-$0.933 &      0.124 &      343.5          & epic \\
 12 & 165313.2$-$415222 &   0.4  &      113.3 &      1.885 &      0.200 &      0.400 &      0.089 &      1.141 &      0.142 &      0.344 &      0.108 &      0.481 &      0.098 &     $-$0.537 &      0.121 &      518.6          & epic \\
 13 & 165313.5$-$415133 &   0.6  &      112.7 &      1.832 &      0.198 &      0.326 &      0.079 &      1.187 &      0.139 &      0.320 &      0.117 &      0.569 &      0.091 &     $-$0.576 &      0.128 &      424.0          & epic \\
 14 & 165315.3$-$415011 &   0.8  &        8.1 &      0.341 &      0.103 &      0.058 &      0.052 &      0.283 &      0.072 &      0.000 &      0.053 &      0.662 &      0.263 &     $-$1.000 &      0.373 &       79.0          & epic \\
\hline
\end{tabular}
\vfill
\end{minipage}
\end{sidewaystable*}

%%%%%%%%%%%%%%%%%%%%%%%%%%%%%%%%%%%%%%%%%%%%%%%%%%%%%%%%%%%%%%%%%%%%%%%%%%%%%%%
%%%%%%%%%%%%%%%%%%%%%%%%%%%%%%%%%%%%%%%%%%%%%%%%%%%%%%%%%%%%%%%%%%%%%%%%%%%%%%%
%%%%%%%%%%%%%%%%%%%%%%%%%%%%%%%%%%%%%%%%%%%%%%%%%%%%%%%%%%%%%%%%%%%%%%%%%%%%%%%

%%%%%%%%%%%%%%%%%%%%%%%%%%%%%%%%%%%%%%%%%%%%%%%%%%%%%%%%%%%%%%%%%%%%%%%%%%%%%%%
%%%%%%%%%%%%%%%%%%%%%%%%% EXTENDED SOURCE CATALOGUE %%%%%%%%%%%%%%%%%%%%%%%%%%%
%%%%%%%%%%%%%%%%%%%%%%%%%%%%%%%%%%%%%%%%%%%%%%%%%%%%%%%%%%%%%%%%%%%%%%%%%%%%%%%

\begin{sidewaystable*}
\begin{minipage}[t][180mm]{\textwidth}
  \caption{Same as Table \ref{tab: Xcat} but for the extended sources. $\sigma_\mathrm{psf}$ (Col. 3b) gives, in arcsec, the $\sigma$ extent of the adjusted Gaussian. }
\label{tab: Xext}
\tiny
\begin{tabular}{c c c c | c c c c c c c c c c c c c | c c}
\hline 
\hline
  \multicolumn{4}{c|}{XMMU J} &\multicolumn{13}{c|}{\epicpn\ instrument}  \\
X\# & HHMMSS.s$\pm$DDAMAS & $\sigma_{\alpha\delta}$ & $\sigma_\mathrm{psf}$ &
$L_2^\mathrm{pn}$ & cr$_\mathrm{pn}$ & $\sigma_\mathrm{pn}$ & 
cr$^\mathrm{S}_\mathrm{pn}$ & $\sigma^\mathrm{S}_\mathrm{pn}$ & 
cr$^\mathrm{M}_\mathrm{pn}$ & $\sigma^\mathrm{M}_\mathrm{pn}$ & 
cr$^\mathrm{H}_\mathrm{pn}$ & $\sigma^\mathrm{H}_\mathrm{pn}$ & 
$HR_1^\mathrm{pn}$  & $\sigma_{HR1}^\mathrm{pn}$ &
$HR_2^\mathrm{pn}$  & $\sigma_{HR2}^\mathrm{pn}$ \\ 

 [1] & [2] & [3a] & [3b] & [4] & [5] & [6] & [7] & [8] & [9] & [10] & [11] & [12] & [13] & [14] & [15] & [16]   \\
\hline
125 & 165351.7$-$414850 & 0.8  & 2.3 &      307.6 &      5.585 &      0.337 &      2.457 &      0.206 &      2.761 &      0.218 &      0.367 &      0.154 &      0.058 &      0.057 &     $-$0.765 &      0.089   \\
133 & 165353.3$-$415101 & 1.42  & 3.1 &       --- &        --- &        --- &        --- &        --- &        --- &        --- &        --- &        --- &        --- &        --- &        --- &        ---   \\
179 & 165359.3$-$415937 & 0.5  & 1.3 &      355.3 &      8.288 &      0.502 &      3.692 &      0.298 &      3.378 &      0.291 &      1.218 &      0.280 &     $-$0.045 &      0.059 &     $-$0.470 &      0.096   \\
204 & 165401.9$-$414335 & 0.5  & 1.4 &       --- &        --- &        --- &        --- &        --- &        --- &        --- &        --- &        --- &        --- &        --- &        --- &        ---   \\
221 & 165404.0$-$415031 & 0.6  & 3.3 &      383.8 &     12.315 &      0.658 &      5.163 &      0.408 &      6.498 &      0.450 &      0.654 &      0.252 &      0.114 &      0.052 &     $-$0.817 &      0.065   \\
229 & 165405.0$-$414234 & 0.5  & 1.7 &      328.1 &      6.858 &      0.399 &      2.519 &      0.230 &      3.421 &      0.249 &      0.918 &      0.210 &      0.152 &      0.057 &     $-$0.577 &      0.080   \\
239 & 165405.6$-$414451 & 0.8  & 2.3 &      124.2 &      4.450 &      0.364 &      2.029 &      0.239 &      2.421 &      0.265 &      0.000 &      0.074 &      0.088 &      0.080 &     $-$1.000 &      0.061   \\
247 & 165406.4$-$414904 & 1.8  & 6.5 &       --- &        --- &        --- &        --- &        --- &        --- &        --- &        --- &        --- &        --- &        --- &        --- &        ---   \\
265 & 165408.5$-$415021 & 0.8  & 4.9 &      540.7 &     20.183 &      0.954 &     11.738 &      0.637 &      8.077 &      0.629 &      0.368 &      0.329 &     $-$0.185 &      0.046 &     $-$0.913 &      0.075   \\
270 & 165409.0$-$415813 & 1.0  & 2.0 &       84.7 &      3.791 &      0.390 &      1.384 &      0.214 &      1.690 &      0.230 &      0.718 &      0.232 &      0.100 &      0.102 &     $-$0.404 &      0.147   \\
281 & 165410.3$-$414840 & 1.0  & 3.2 &       --- &        --- &        --- &        --- &        --- &        --- &        --- &        --- &        --- &        --- &        --- &        --- &        ---   \\
289 & 165411.1$-$415233 & 0.6  & 2.9 &      520.0 &     10.391 &      0.461 &      4.518 &      0.286 &      5.270 &      0.307 &      0.604 &      0.190 &      0.077 &      0.043 &     $-$0.794 &      0.059   \\
344 & 165417.0$-$414540 & 0.5  & 1.9 &      385.5 &      6.539 &      0.354 &      2.457 &      0.209 &      3.041 &      0.225 &      1.041 &      0.175 &      0.106 &      0.056 &     $-$0.490 &      0.070   \\
377 & 165420.2$-$414849 & 0.6  & 2.1 &       --- &        --- &        --- &        --- &        --- &        --- &        --- &        --- &        --- &        --- &        --- &        --- &        ---   \\
383 & 165421.1$-$415706 & 1.5  & 2.8 &       --- &        --- &        --- &        --- &        --- &        --- &        --- &        --- &        --- &        --- &        --- &        --- &        ---   \\
411 & 165424.4$-$414935 & 0.8  & 3.2 &      356.6 &     10.426 &      0.537 &      3.783 &      0.331 &      5.623 &      0.354 &      1.019 &      0.232 &      0.196 &      0.052 &     $-$0.693 &      0.061   \\
417 & 165425.5$-$415424 & 0.7  & 1.5 &       --- &        --- &        --- &        --- &        --- &        --- &        --- &        --- &        --- &        --- &        --- &        --- &        ---   \\
451 & 165429.6$-$414850 & 0.8  & 2.8 &      316.7 &      7.933 &      0.474 &      2.578 &      0.263 &      4.399 &      0.321 &      0.955 &      0.228 &      0.261 &      0.058 &     $-$0.643 &      0.073   \\
456 & 165429.9$-$413921 & 1.3  & 2.3 &       --- &        --- &        --- &        --- &        --- &        --- &        --- &        --- &        --- &        --- &        --- &        --- &        ---   \\
%542 & 165443.1$-$415634 & 0.7  & NULL&       17.1 &      2.546 &      0.345 &      0.878 &      0.165 &      1.073 &      0.204 &      0.595 &      0.224 &      0.100 &      0.133 &     $-$0.287 &      0.193   \\
\hline
\vspace*{0mm}\\
\hline
\hline
  \multicolumn{4}{c|}{XMMU J} & \multicolumn{13}{c|}{\epicmos1 instrument} \\
X\# & HHMMSS.s$\pm$DDAMAS & $\sigma_{\alpha\delta}$ & $\sigma_\mathrm{psf}$ &
$L_2^\mathrm{MOS1}$ & cr$_\mathrm{MOS1}$ & $\sigma_\mathrm{MOS1}$ & 
cr$^\mathrm{S}_\mathrm{MOS1}$ & $\sigma^\mathrm{S}_\mathrm{MOS1}$ & 
cr$^\mathrm{M}_\mathrm{MOS1}$ & $\sigma^\mathrm{M}_\mathrm{MOS1}$ & 
cr$^\mathrm{H}_\mathrm{MOS1}$ & $\sigma^\mathrm{H}_\mathrm{MOS1}$ & 
$HR_1^\mathrm{MOS1}$  & $\sigma_{HR1}^\mathrm{MOS1}$ &
$HR_2^\mathrm{MOS1}$  & $\sigma_{HR2}^\mathrm{MOS1}$ \\

[1] & [2] & [3a] & [3b] & [17] &[18] & [19] & [20] & [21] & [22] & [23] & [24] & [25] & [26] & [27] & [28] & [29] \\
\hline
125 & 165351.7$-$414850 & 0.8  & 2.3 &    72.7 &      1.397 &      0.155 &      0.468 &      0.082 &      0.824 &      0.110 &      0.105 &      0.071 &      0.276 &      0.102 &     $-$0.774 &      0.139  \\
133 & 165353.3$-$415101 & 1.4  & 3.1 &    61.4 &      1.483 &      0.181 &      0.458 &      0.092 &      0.930 &      0.131 &      0.096 &      0.086 &      0.340 &      0.108 &     $-$0.812 &      0.154  \\
179 & 165359.3$-$415937 & 0.5  & 1.3 &   168.5 &      2.799 &      0.241 &      0.900 &      0.121 &      1.398 &      0.157 &      0.501 &      0.138 &      0.217 &      0.083 &     $-$0.472 &      0.116  \\
204 & 165401.9$-$414335 & 0.5  & 1.4 &   961.2 &      6.975 &      0.383 &      1.679 &      0.202 &      4.103 &      0.281 &      1.193 &      0.166 &      0.419 &      0.057 &     $-$0.549 &      0.054  \\
221 & 165404.0$-$415031 & 0.6  & 3.3 &   135.1 &      3.762 &      0.316 &      1.065 &      0.167 &      2.275 &      0.230 &      0.421 &      0.137 &      0.362 &      0.081 &     $-$0.688 &      0.090  \\
229 & 165405.0$-$414234 & 0.5  & 1.7 &   136.3 &      2.204 &      0.196 &      0.629 &      0.099 &      1.317 &      0.138 &      0.258 &      0.098 &      0.354 &      0.083 &     $-$0.672 &      0.108  \\
239 & 165405.6$-$414451 & 0.8  & 2.3 &    36.8 &      1.355 &      0.192 &      0.334 &      0.094 &      0.888 &      0.139 &      0.134 &      0.094 &      0.454 &      0.128 &     $-$0.738 &      0.163  \\
247 & 165406.4$-$414904 & 1.8  & 6.5 &    48.0 &      5.111 &      0.876 &      2.860 &      0.763 &      2.250 &      0.415 &      0.000 &      0.111 &     $-$0.119 &      0.160 &     $-$1.000 &      0.099  \\
265 & 165408.5$-$415021 & 0.8  & 4.9 &   171.9 &      5.746 &      0.410 &      2.657 &      0.273 &      3.089 &      0.302 &      0.000 &      0.047 &      0.075 &      0.071 &     $-$1.000 &      0.030  \\
270 & 165409.0$-$415813 & 1.0  & 2.0 &    46.9 &      1.368 &      0.192 &      0.276 &      0.078 &      0.871 &      0.133 &      0.221 &      0.115 &      0.518 &      0.117 &     $-$0.595 &      0.174  \\
281 & 165410.3$-$414840 & 1.0  & 3.2 &   215.9 &      6.795 &      0.536 &      1.987 &      0.258 &      4.223 &      0.337 &      0.584 &      0.328 &      0.360 &      0.066 &     $-$0.757 &      0.121  \\
289 & 165411.1$-$415233 & 0.6  & 2.9 &   220.8 &      3.566 &      0.243 &      0.747 &      0.114 &      2.369 &      0.182 &      0.449 &      0.113 &      0.521 &      0.062 &     $-$0.681 &      0.071  \\
344 & 165417.0$-$414540 & 0.5  & 1.9 &    93.0 &      1.812 &      0.175 &      0.472 &      0.086 &      0.980 &      0.122 &      0.359 &      0.091 &      0.350 &      0.097 &     $-$0.464 &      0.111  \\
377 & 165420.2$-$414849 & 0.6  & 2.1 &   375.0 &      5.208 &      0.272 &      1.112 &      0.127 &      3.227 &      0.203 &      0.870 &      0.130 &      0.487 &      0.050 &     $-$0.575 &      0.054  \\
383 & 165421.1$-$415706 & 1.5  & 2.8 &    71.3 &      2.593 &      0.288 &      0.426 &      0.100 &      1.537 &      0.206 &      0.630 &      0.175 &      0.566 &      0.092 &    $-$0.419 &      0.127  \\
411 & 165424.4$-$414935 & 0.8  & 3.2 &    95.4 &      3.133 &      0.283 &      0.730 &      0.153 &      2.078 &      0.209 &      0.325 &      0.113 &      0.480 &      0.089 &     $-$0.730 &      0.085  \\
417 & 165425.5$-$415424 & 0.7  & 1.5 &    89.1 &      2.213 &      0.228 &      0.482 &      0.102 &      1.374 &      0.170 &      0.357 &      0.112 &      0.481 &      0.094 &     $-$0.588 &      0.110  \\
451 & 165429.6$-$414850 & 0.8  & 2.8 &    80.1 &      2.315 &      0.246 &      0.615 &      0.128 &      1.562 &      0.183 &      0.138 &      0.104 &      0.435 &      0.097 &     $-$0.837 &      0.114  \\
456 & 165429.9$-$413921 & 1.3  & 2.3 &    68.6 &      2.927 &      0.366 &      0.527 &      0.119 &      1.906 &      0.260 &      0.493 &      0.228 &      0.567 &      0.089 &     $-$0.589 &      0.157  \\
%542 & 165443.1$-$415634 & 0.7  & NULL&     7.6 &      1.244 &      0.215 &      0.201 &      0.084 &      0.815 &      0.149 &      0.228 &      0.131 &      0.605 &      0.144 &     $-$0.563 &      0.206  \\
\hline
\vspace*{0mm}\\
\hline
\hline
  \multicolumn{4}{c|}{XMMU J} & \multicolumn{13}{c|}{\epicmos2 instrument}  \\
X\# & HHMMSS.s$\pm$DDAMAS & $\sigma_{\alpha\delta}$ &  $\sigma_\mathrm{psf}$ &
$L_2^\mathrm{MOS2}$ & cr$_\mathrm{MOS2}$ & $\sigma_\mathrm{MOS2}$ & 
cr$^\mathrm{S}_\mathrm{MOS2}$ & $\sigma^\mathrm{S}_\mathrm{MOS2}$ & 
cr$^\mathrm{M}_\mathrm{MOS2}$ & $\sigma^\mathrm{M}_\mathrm{MOS2}$ & 
cr$^\mathrm{H}_\mathrm{MOS2}$ & $\sigma^\mathrm{H}_\mathrm{MOS2}$ & 
$HR_1^\mathrm{MOS2}$  & $\sigma_{HR1}^\mathrm{MOS2}$ &
$HR_2^\mathrm{MOS2}$  & $\sigma_{HR2}^\mathrm{MOS2}$ &
$L_2^\mathrm{EPIC}$ & Instr.  \\ 

[1] & [2] & [3a] & [3b]  & [30]  & [31] & [32] & [33] & [34] & [35] & [36] & [37] & [38] & [39] & [40] & [41] & [42] & [43] & [44]   \\
\hline
125 & 165351.7$-$414850 & 0.8  & 2.3 &    111.1 &      1.775 &      0.162 &      0.539 &      0.086 &      1.025 &      0.111 &      0.211 &      0.081 &      0.311 &      0.087 &     $-$0.659 &      0.113 &      492.7 & epic   \\
133 & 165353.3$-$415101 & 1.4  & 3.1 &    127.0 &      2.256 &      0.211 &      0.801 &      0.120 &      1.311 &      0.147 &      0.144 &      0.093 &      0.241 &      0.088 &     $-$0.802 &      0.116 &      189.1 & mos    \\
179 & 165359.3$-$415937 & 0.5  & 1.3 &    135.8 &      2.387 &      0.225 &      0.908 &      0.118 &      1.033 &      0.140 &      0.446 &      0.130 &      0.064 &      0.094 &     $-$0.397 &      0.136 &      661.8 & epic   \\
204 & 165401.9$-$414335 & 0.5  & 1.4 &     --- &        --- &        --- &        --- &        --- &        --- &        --- &        --- &        --- &        --- &        --- &        --- &        --- &      961.2 & m1     \\
221 & 165404.0$-$415031 & 0.6  & 3.3 &    149.4 &      3.477 &      0.297 &      0.914 &      0.165 &      2.517 &      0.235 &      0.046 &      0.078 &      0.467 &      0.079 &     $-$0.964 &      0.060 &      670.4 & epic   \\
229 & 165405.0$-$414234 & 0.5  & 1.7 &    126.2 &      2.106 &      0.188 &      0.684 &      0.103 &      1.171 &      0.131 &      0.251 &      0.087 &      0.262 &      0.087 &     $-$0.647 &      0.106 &      592.5 & epic   \\
239 & 165405.6$-$414451 & 0.8  & 2.3 &     32.1 &      1.287 &      0.189 &      0.454 &      0.097 &      0.659 &      0.130 &      0.174 &      0.097 &      0.184 &      0.140 &     $-$0.583 &      0.195 &      193.7 & epic   \\
247 & 165406.4$-$414904 & 1.8  & 6.5 &     36.5 &      4.491 &      0.565 &      2.123 &      0.374 &      2.341 &      0.403 &      0.026 &      0.131 &      0.049 &      0.123 &     $-$0.978 &      0.109 &       85.1 & mos    \\
265 & 165408.5$-$415021 & 0.8  & 4.9 &    198.5 &      6.099 &      0.420 &      2.680 &      0.278 &      3.410 &      0.305 &      0.010 &      0.082 &      0.120 &      0.067 &     $-$0.994 &      0.048 &      913.3 & epic   \\
270 & 165409.0$-$415813 & 1.0  & 2.0 &     49.3 &      1.589 &      0.209 &      0.365 &      0.091 &      0.837 &      0.139 &      0.387 &      0.127 &      0.393 &      0.127 &     $-$0.368 &      0.159 &      182.2 & epic   \\
281 & 165410.3$-$414840 & 1.0  & 3.2 &    104.4 &      4.584 &      0.415 &      1.208 &      0.234 &      2.987 &      0.308 &      0.390 &      0.151 &      0.424 &      0.090 &     $-$0.769 &      0.082 &      321.3 & mos    \\
289 & 165411.1$-$415233 & 0.6  & 2.9 &    224.9 &      3.470 &      0.236 &      0.963 &      0.126 &      2.272 &      0.173 &      0.234 &      0.100 &      0.405 &      0.063 &     $-$0.813 &      0.074 &      968.4 & epic   \\
344 & 165417.0$-$414540 & 0.5  & 1.9 &    102.5 &      1.697 &      0.169 &      0.455 &      0.084 &      1.131 &      0.128 &      0.112 &      0.072 &      0.426 &      0.088 &     $-$0.820 &      0.107 &      582.4 & epic   \\
377 & 165420.2$-$414849 & 0.6  & 2.1 &    383.3 &      5.205 &      0.272 &      1.056 &      0.123 &      3.280 &      0.202 &      0.870 &      0.134 &      0.513 &      0.049 &     $-$0.581 &      0.055 &      760.0 & mos    \\
383 & 165421.1$-$415706 & 1.5  & 2.8 &     89.9 &      3.020 &      0.309 &      0.514 &      0.127 &      1.985 &      0.227 &      0.521 &      0.166 &      0.588 &      0.089 &     $-$0.584 &      0.111 &      162.1 & mos    \\
411 & 165424.4$-$414935 & 0.8  & 3.2 &     84.6 &      3.007 &      0.282 &      0.860 &      0.154 &      1.831 &      0.203 &      0.316 &      0.121 &      0.361 &      0.092 &     $-$0.706 &      0.100 &      537.7 & epic   \\
417 & 165425.5$-$415424 & 0.7  & 1.5 &     82.4 &      2.308 &      0.242 &      0.494 &      0.109 &      1.268 &      0.170 &      0.545 &      0.133 &      0.439 &      0.104 &     $-$0.399 &      0.117 &      172.4 & mos    \\
451 & 165429.6$-$414850 & 0.8  & 2.8 &     68.7 &      2.330 &      0.251 &      0.560 &      0.123 &      1.353 &      0.180 &      0.417 &      0.123 &      0.415 &      0.106 &     $-$0.529 &      0.117 &      466.5 & epic   \\
456 & 165429.9$-$413921 & 1.3  & 2.3 &     78.8 &      2.680 &      0.335 &      0.941 &      0.167 &      1.480 &      0.224 &      0.260 &      0.186 &      0.223 &      0.111 &     $-$0.701 &      0.186 &      148.3 & mos    \\
%542 & 165443.1$-$415634 & 0.7  & NULL&     20.5 &      1.626 &      0.234 &      0.223 &      0.094 &      1.113 &      0.174 &      0.291 &      0.126 &      0.666 &      0.125 &     $-$0.586 &      0.152 &       45.3 & epic   \\
\hline
\end{tabular}
\vfill
\end{minipage}
\end{sidewaystable*}
%%%%%%%%%%%%%%%%%%%%%%%%%%%%%%%%%%%%%%%%%%%%%%%%%%%%%%%%%%%%%%%%%%%%%%%%%%%%%%%
%%%%%%%%%%%%%%%%%%%%%%%%%%%%%%%%%%%%%%%%%%%%%%%%%%%%%%%%%%%%%%%%%%%%%%%%%%%%%%%
%%%%%%%%%%%%%%%%%%%%%%%%%%%%%%%%%%%%%%%%%%%%%%%%%%%%%%%%%%%%%%%%%%%%%%%%%%%%%%%

The described detection procedure was applied for each \epic\ instrument as well as for any combination of them. The resulting source lists were generally consistent. The main difference comes from the presence of different gaps in the different data sets. We built our final source list  adopting the following criteria. 

(i) We selected the deepest combination of \epic\ instruments, requiring that the detected source is distant by at least 13\arcsec\ from any gap, bad column or detector edge.

(ii) By a visual inspection of each source in all images and subsequent combinations, we adopted an empirical equivalent logarithmic likelihood ($L_2$) threshold as the detection limit. This led us to consider the way to perform a consistent choice for the threshold values while dealing with different instrument combinations, and hence with different numbers of input images. 
As a general comment, it is obvious that adopting a constant logarithmic likelihood threshold while dealing with different combinations of the \epic\ instruments does not allow us to keep a constant threshold in terms of the signal level. Indeed, all other things being equal, the signal-to-noise ratio is increased while combining several detectors, allowing us in principle to detect fainter sources. However, in such a crowded field as the current one, we note that no important gain is achieved in terms of source detection. In other words, the very large majority of the detected sources are already seen in a single instrument, though of course combining the different instruments yields a much better estimation of their X-ray parameters. 

As a consequence, we have decided to adopt a logarithmic likelihood threshold in one instrument and to look for the  equivalent thresholds in any \epic\ combination.
 This issue is presented into more details in Appendix \ref{app: L2_b}. Table \ref{tab: l2} gives the  logarithmic likelihood thresholds finally used for the source detection. These values provide thresholds in various combinations that are consistent with the logarithmic likelihood-based detection threshold adopted in a single \mos\ instrument. 
We note that this procedure does not modify the spatial response of the detectors and that  the known variations of the \epic\ instrument sensitivity with the axial distance will of course still affect our results.

%------------------------------ FINDING CHARTS   ---------------------------------
   \begin{figure*}
   \centering
   \includegraphics[width=\textwidth]{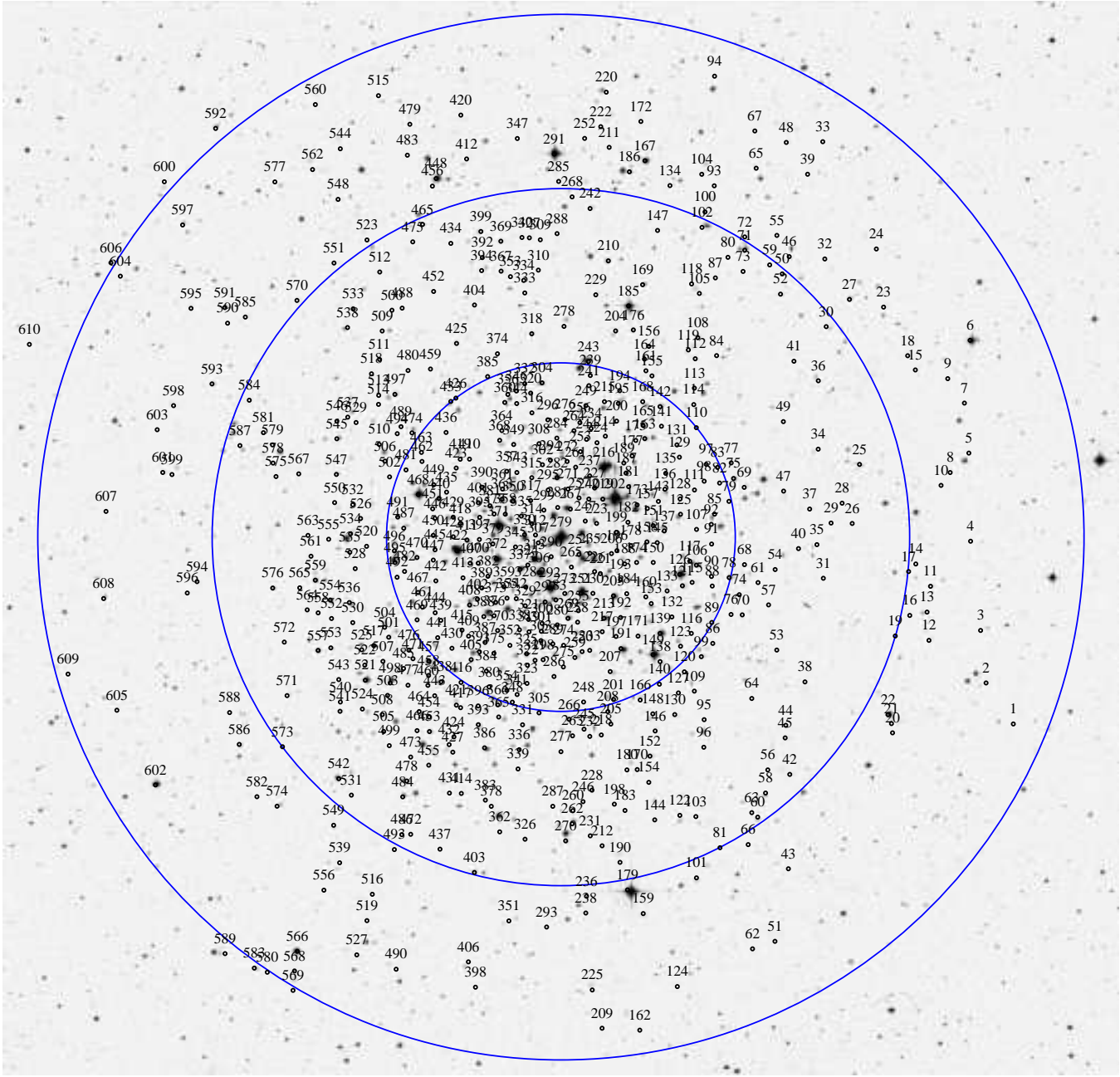}
   \caption{Detected X-ray sources overlaid on a DSS1 image ($V$ band) of the \xmm\ FOV. The sources are indicated by black circles, with a radius of 3\arcsec, similar to the adopted cross-correlation radius. The numbers above these circles give the internal X-ray  source identification as provided in the first column of Table \ref{tab: Xcat}.  The three blue circles indicate regions with a radius of 5\arcmin, 10\arcmin\ and 15\arcmin\ around X\#\,279 (\hda). North is up and East is to the left.}
         \label{fig: dss_15}
   \end{figure*}

   \begin{figure*}
   \centering
   \includegraphics[width=\textwidth]{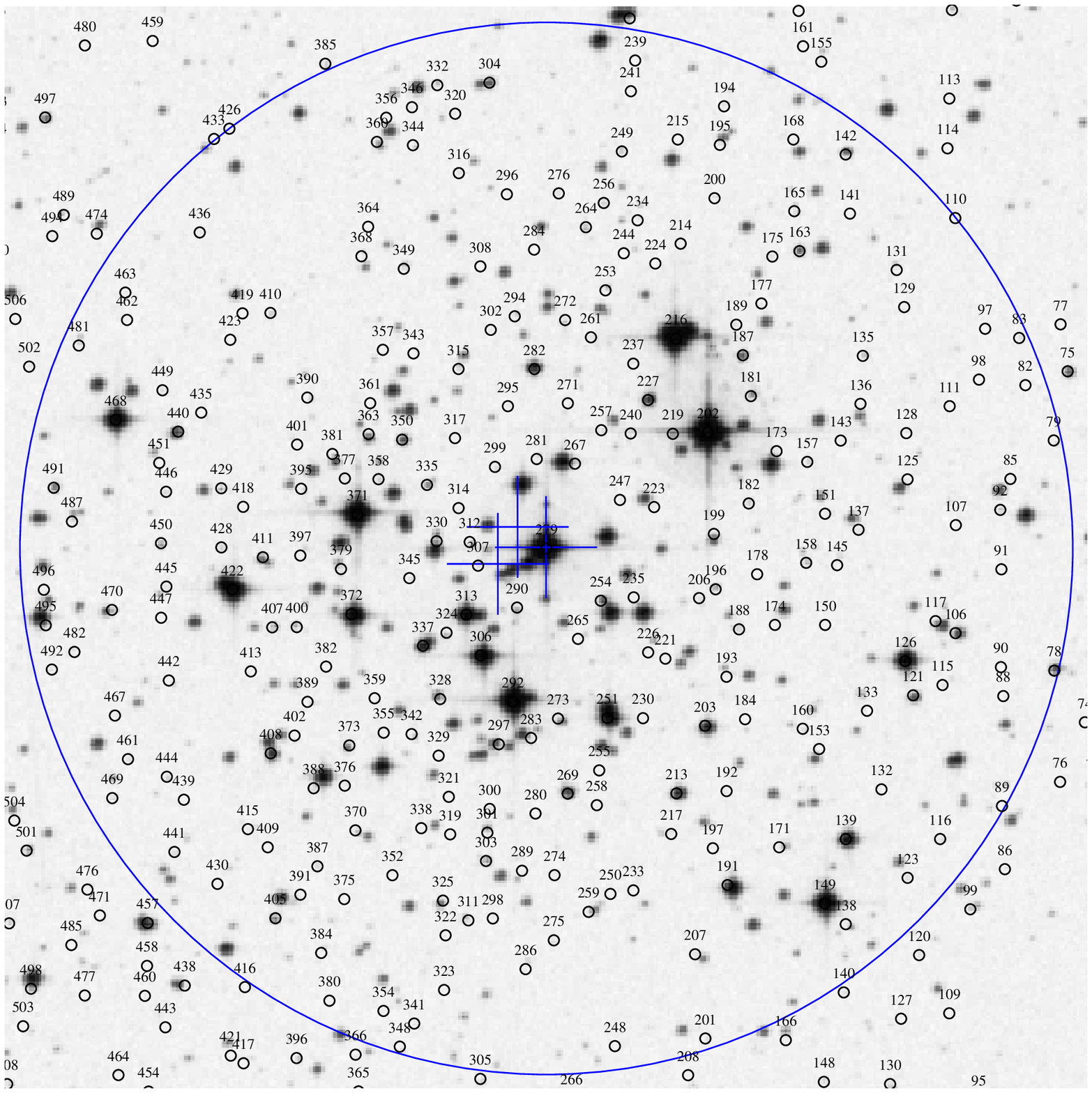}
   \caption{Same as Fig.~\ref{fig: dss_15}, zoomed on the inner part of the FOV. The circle radius is 5\arcmin. From left to right, the three crosses respectively give the position of the geometrical centre of the cluster, its X-ray emission centre (computed adopting the {\it pn-equivalent} count rates for each source), and the position of \hda\ (X\# 279).}
         \label{fig: dss_5}
   \end{figure*}
%---------------------------------------------------------------------------------

(iii) In the few cases for which multi-source fitting was relevant, we adopted the results obtained with this fitting. We however paid a special attention to reject cases of fake multi-fitting sometimes induced by near-gap/edge effects or by multiple entries for a unique X-ray source in the preliminary source list. 

(iv) We finally checked every source in the final list by individually looking at the different image combinations. We eliminated the very few double entries in the list. Doing this, we noticed a couple of presumably physical sources that were ignored by the detection algorithm. We decided to include those sources in the input source list of the {\it emldetect} task. Most of them were satisfactorily fitted, giving an equivalent logarithmic likelihood above the adopted detection threshold. These additional sources were included in the final catalogue.

(v) The main X-ray catalogue presented in Table \ref{tab: Xcat} is based on the  point-like source detection only. For some sources, the equivalent logarithmic likelihood $L_2$ is significantly increased if one adjusts an extended source model rather than a point-like model. These sources are flagged in Table \ref{tab: Xcat} and we provide, in Table \ref{tab: Xext}, a complementary extended source catalogue that gives, in addition to the results listed in the  main catalogue,  the {\it emldetect} extended-psf fit  results for these sources. \\

The final catalogue  lists 610 sources in the \xmm\ FOV, among which 19 are flagged as extended. Based on the {\it edetect\_chain} results, it provides, among other information, the source position, the total count rates in the different instruments and the two hardness ratios :
\begin{eqnarray}
 HR_1=\frac{M_\mathrm{X}-S_\mathrm{X}}{M_\mathrm{X}+S_\mathrm{X}} \vspace*{2mm}, \label{eq: hr1}\\
 HR_2=\frac{H_\mathrm{X}-M_\mathrm{X}}{H_\mathrm{X}+M_\mathrm{X}} \vspace*{2mm}. \label{eq: hr2}
\end{eqnarray}
A sample of the catalogue is provided in Table \ref{tab: Xcat} while Table \ref{tab: Xext} gives the  complementary catalogue for the 19 extended X-ray sources detected. In addition, source X\#234 appears clearly double in the \epic\ image though it is not detected as an extended object. Table \ref{tab: Xcat} is available online via the Centre de Donn\'ees astronomiques de Strasbourg (CDS, http://cdsweb.u-strasbg.fr). Finding charts for the X-ray sources are provided by Figs. \ref{fig: dss_15} and \ref{fig: dss_5}.

%---------------------------- JEFFRIES BEST_FIT PARAMETERS ----------------------------------------------------------------

\begin{table*}
\centering
\caption{Best fit parameters (Cols. 3-5) of the $\Phi(r)$ function (Eq.~\ref{eq: jtp97}) for different optical/infrared catalogues. The catalogue name is given in Col.~1 along with the relevant number of X-ray sources in the corresponding field (Col.~2). The adopted correlation radius ($r_\mathrm{corr}$) for identification is given in Col. 6. Col. 7 lists the actual number $N_\mathrm{corr}$ of identified X-ray sources (see Table \ref{tab: Xcat_iden}) and the corresponding percentage related to the considered number of X-ray sources ($N_\mathrm{X}$). Col. 8 gives the number of associated counterparts predicted by the distribution $\Phi(r)$ (Eq.~\ref{eq: jtp97}) at a radius equal to $r_\mathrm{corr}$. It also provides the corresponding percentage of theoretically identified X-ray sources. The next two columns provide, among the  number of associated counterparts $\Phi(r_\mathrm{corr})$, the predicted number of true ($\Phi_\mathrm{true}(r_\mathrm{corr})$) and spurious ($\Phi_\mathrm{spur.}(r_\mathrm{corr})$) counterparts. The contribution  of true and spurious counterparts to the total number of (theoretically) associated optical sources are also given in the corresponding columns. }
\label{tab: jtp97}
\centering
\begin{tabular}{c c c c c c c c c c}
\hline
\hline
Opt. Cat. & $N_\mathrm{X}$ & $A$ & $\sigma$             & $B$                          & $r_\mathrm{corr}$ & $N_\mathrm{corr}$ & $\Phi(r_\mathrm{corr})$ & $\Phi_\mathrm{true}(r_\mathrm{corr}$) & $\Phi_\mathrm{spur.}(r_\mathrm{corr}$) \\
    &           &     & (\arcsec) & $10^{-3} ($\arcsec$)^{-2}$ & (\arcsec) \\
\hline
2MASS        & 610 & 322.2 & 0.91 & 30.0 & 2 & 384 (63.0\%) & 383.8 (62.9\%) & 293.3 (76.4\%) & 90.5 (23.6\%) \\
GSC~2.2      & 610 & 384.3 & 1.25 & 3.5  & 3 & 372 (61.0\%) & 383.7 (62.9\%) & 362.7 (94.5\%) & 21.1 (5.5\%)  \\
USNO         & 610 & 383.1 & 1.24 & 0.8  & 3 & 344 (56.4\%) & 367.8 (60.3\%) & 362.8 (98.6\%) &  5.0 (1.4\%)  \\
SBL98~v2     & 536 & 431.8 & 1.09 & 11.9 & 3 & 447 (83.4\%) & 451.9 (84.3\%) & 422.1 (93.4\%) & 29.8 (6.6\%)  \\
\sbl: $V<19$ & 610 & 396.6 & 1.01 & 4.3  &2.5& 384 (63.0\%) & 395.7 (64.9\%) & 378.4 (95.6\%) & 17.3 (4.4\%) \\
\sbl: $V<20$ & 610 & 422.6 & 0.95 & 13.9 &2.5& 450 (73.8\%) & 453.7 (74.4\%) & 408.9 (90.1\%) & 44.9 (9.9\%) \\
\hline
\end{tabular}
\end{table*}

%--------------------------------------------------------------------------------------------------------------------------

\subsection{Source Identification }\label{ssect: source_id}

To determine the optical counterparts of the detected X-ray sources, we cross-correlated our source list with several existing optical/infrared catalogues. We used the  US Naval Observatory \citep[ USNO B1.0]{USNOB10_mnras}, the 2MASS All Sky Data Release \citep{2MASS_mnras} and the Guide Star Catalogue-II \citep{GSC22}. We also made use of the optical catalogue of \citet[ SBL98 hereafter]{SBL98}. However, the star positions in the \citetalias{SBL98} catalogue as available from the Centre de Donn\'ees astronomiques de Strasbourg (CDS) show clear shifts compared to the true positions on the sky. This results from an excessive rounding of the star coordinates in the CDS database: they are given with a precision of respectively one second and one tenth of arcmin on the right ascension and declination. This is far insufficient in such a crowded field as \ngc. We therefore used the original \citetalias{SBL98} catalogue, obtained from the authors and that lists object coordinates a hundred times more precisely. Beyond the 860 objects with $V \le 16$ listed in \citetalias{SBL98}, this UBV(RI)$_\mathrm{C}$~\&~\halph\ catalogue was completed with 7199 objects, extending the first version of the \citetalias{SBL98} catalogue down to $V=21$. However, the \citetalias{SBL98} field of view was limited to a 20\arcmin\ $\times$ 20\arcmin\ area and thus does not cover the whole \epic\ FOV. It can thus not be used for identification throughout the entire field and we selected the X-ray sources that are located in the sub-region of the FOV that is covered by the SBL98~v2 catalogue. This yields a number of X-ray sources $N_\mathrm{X}^\mathrm{SBL}=536$ as quoted in Table \ref{tab: jtp97}. More recently, one of us (H. Sung) acquired new UBV(RI)$_\mathrm{C}$ observations covering a field of about 40\arcmin\ $\times$ 40\arcmin\ around \ngc. 30866 stars were observed down to $V<22$. These observations will be presented in a forthcoming paper (Sung et al.~2006, in preparation) and we only focused here on the resulting catalogue. We will refer to this new catalogue as the \sbl.

  \begin{figure}
   \centering
   \includegraphics[width=4.1cm]{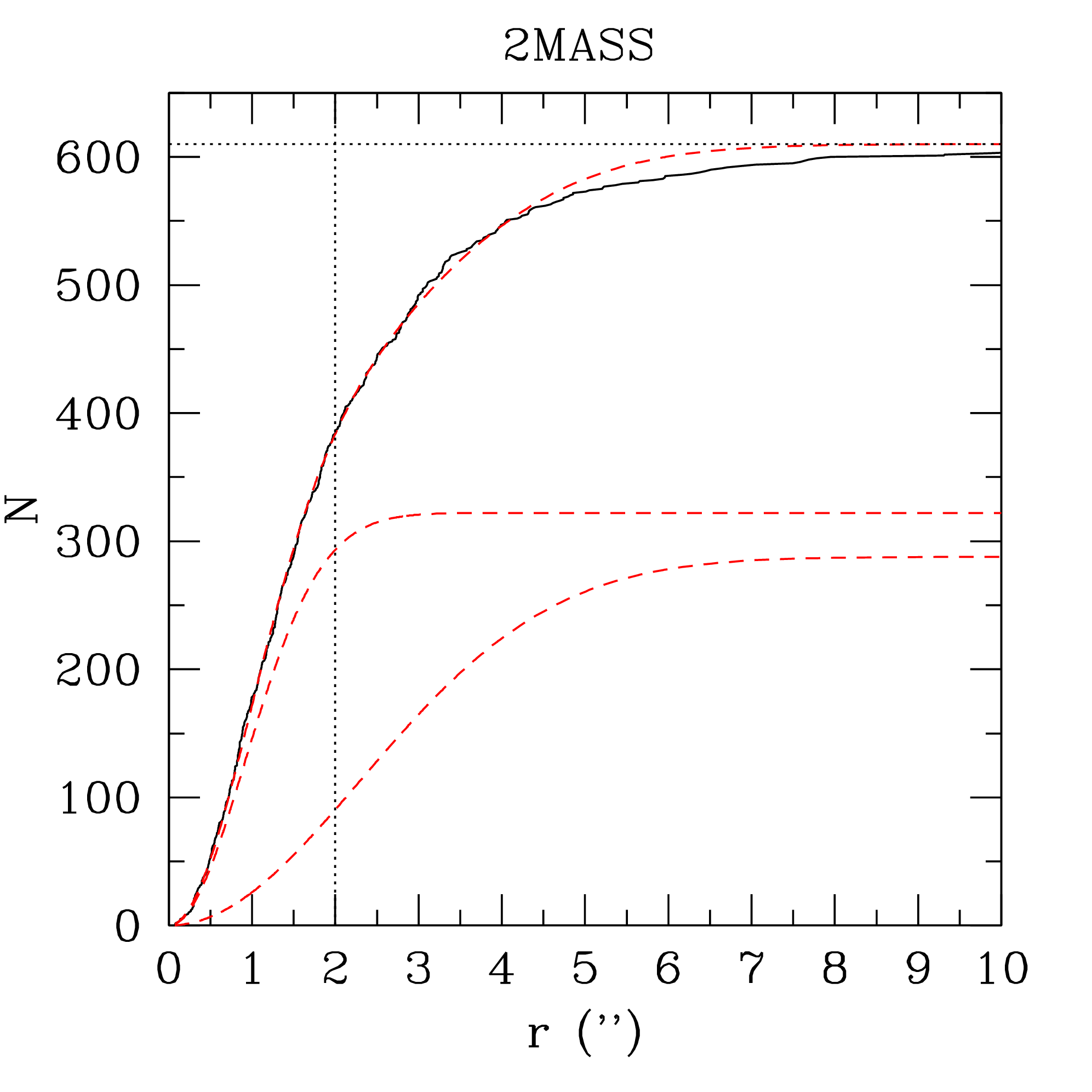}
   \includegraphics[width=4.1cm]{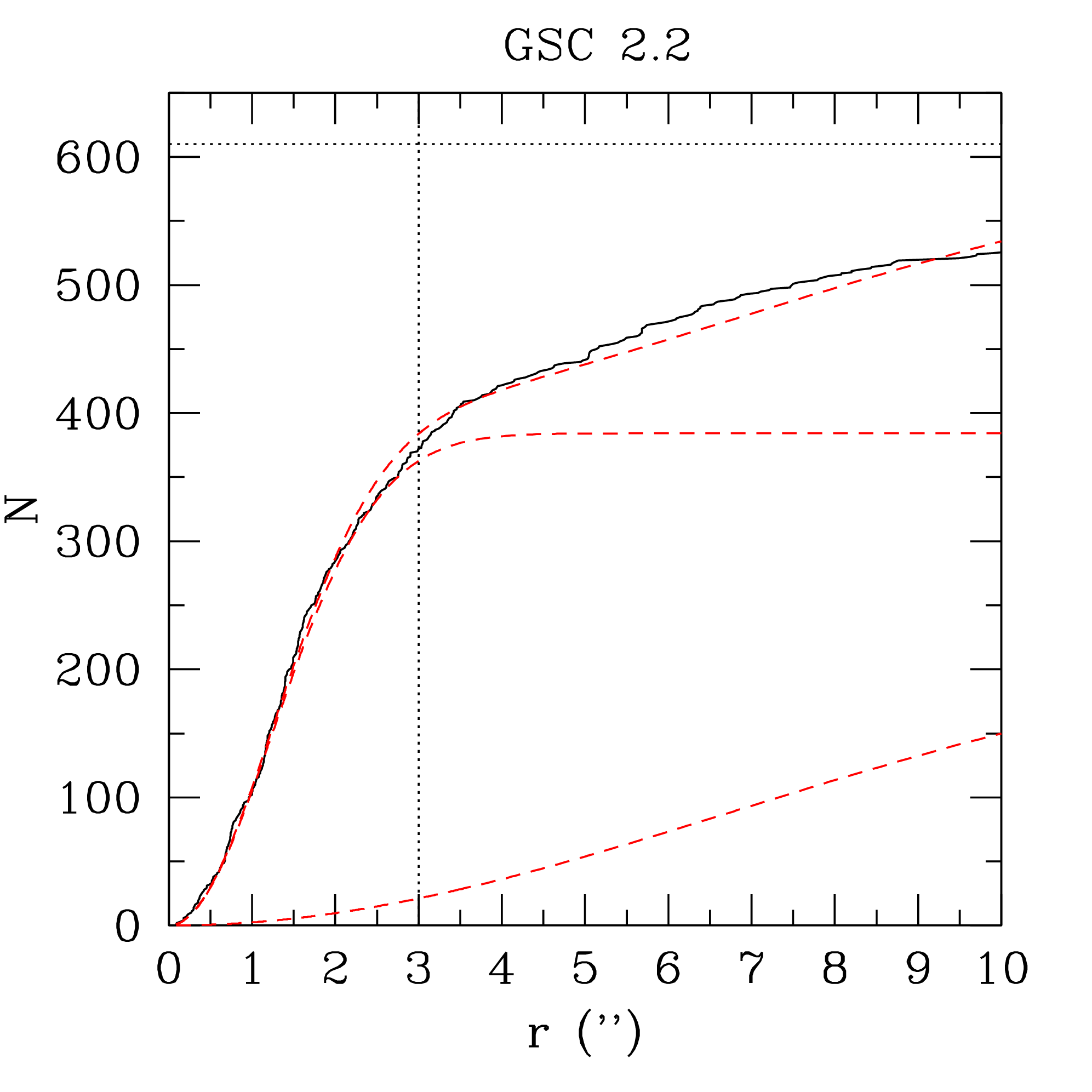}
   \includegraphics[width=4.1cm]{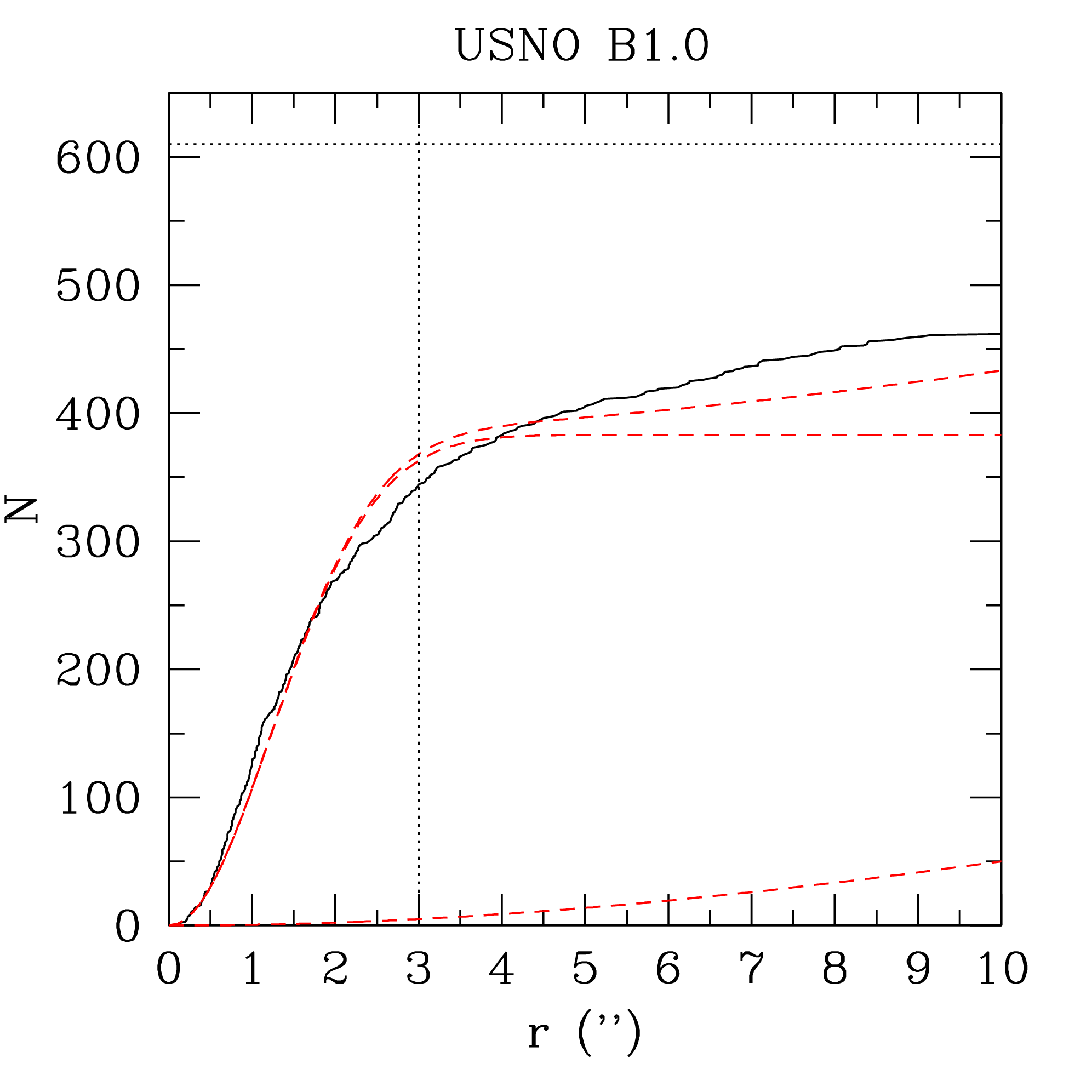}
   \includegraphics[width=4.1cm]{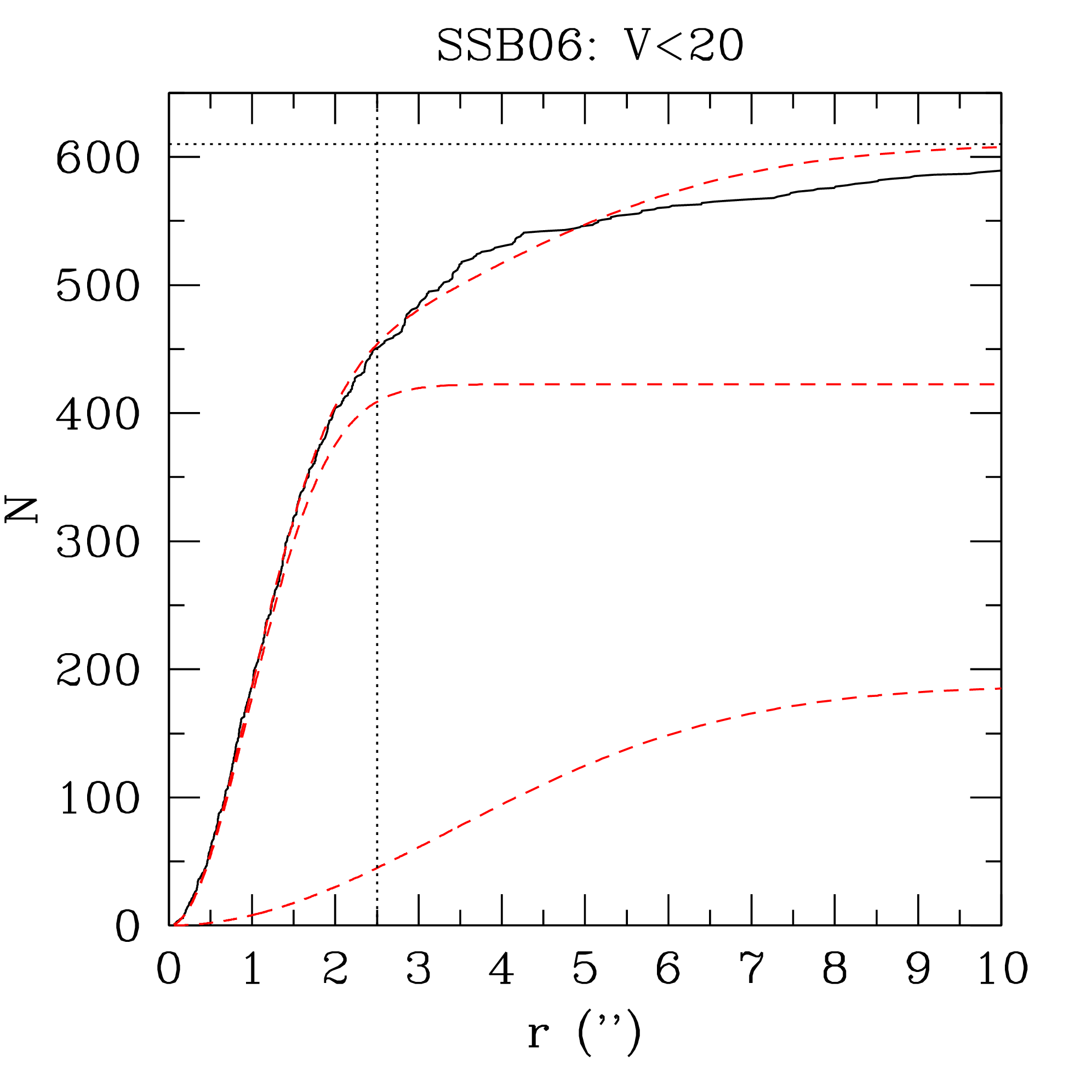}
   \caption{Cumulative distributions (solid lines) of the number ($N$) of closest associated counterparts as a function of the separation radius ($r$) and for the different catalogues used. The horizontal dotted lines show the number $N_\mathrm{X}$ of X-ray sources in the catalogue. The dashed lines, from top to bottom in each panel, correspond respectively to the best-fit $\Phi(r)$ function, to the number of truly associated counterparts $\Phi_\mathrm{true}$ and to the number of spurious ones $\Phi_\mathrm{spur.}$, as a function of the correlation radius $r$. Dotted vertical lines show the correlation radius adopted for the purpose of source identification.
             }
   \label{fig: jtp97}
   \end{figure}

As a first approach, we investigated the possibility of systematic differences between the  reference frames of the different catalogues. For this purpose, we selected the bright O-type stars in the different source lists and we compared their locations to the ones of their X-ray counterparts. Neither a significant systematic shift nor a field rotation was apparent. Typical 1-$\sigma$ dispersions computed on the differences between the locations of the X-ray sources and their optical counterparts are about 0.9\arcsec\ in right ascension and 0.7\arcsec\ in declination. Similarly, the 1-$\sigma$ dispersion on the field rotation is about 3\arcmin.
As a second step and for each of the previously cited catalogues, we determined the closest optical counterpart of each X-ray source in the field of view. We then calculated the cumulative distribution ($\Phi(r)$) of the closest associated counterparts as a function of the individual correlation radius \citep[see][]{JTP97}. The generated diagrams are shown in Fig.~\ref{fig: jtp97}. Following \citeauthor{JTP97}, we assumed that $\Phi(r)$ is formed by two terms: the cumulative distribution of true correlations $\Phi_\mathrm{true}$ and the cumulative number of spurious associations $\Phi_\mathrm{spur.}$. This is expressed in the simple relation: 
\begin{eqnarray}
\Phi(r)&=&\Phi_\mathrm{true}+\Phi_\mathrm{spur.}\\
&=&A \left[ 1- \exp \left( \frac{-r^2}{2\sigma^2} \right) \right]+ \left( N_\mathrm{X}-A \right) \left[ 1 - \exp \left( -\pi r^2 B \right) \right]
\label{eq: jtp97}
\end{eqnarray}
that can be adjusted to the empirical distribution. In Eq.~\ref{eq: jtp97}, $N_\mathrm{X}$ is the number of X-ray sources while $A$ is the number of true correlations with the optical/infrared catalogue.  $B$ is the optical/infrared catalogue density and $\sigma$ is related to the statistical uncertainty on the X-ray source position. 
Though Eq.~\ref{eq: jtp97} is approximative and rests on the hypothesis of a uniform optical population (i.e.\ constant $B$ and constant psf over the full FOV), it fits reasonably well the rising branch of the different curves plotted in Fig.~\ref{fig: jtp97}. Table \ref{tab: jtp97} gives the  obtained values of the $A$, $B$ and $\sigma$ parameters. As the hypothesis of constant $B$ throughout the FOV is clearly violated in the case of \ngc, we also estimated the number of spurious associations using a more empirical approach. We arbitrarily shifted the X-ray source positions by 30\arcsec\ in any given direction and we re-ran the cross-correlation at a fixed $r_\mathrm{corr}$ (either 2\farcs5 or 3\farcs0 according to the catalogue considered). The obtained number of spurious associations is never larger  by more than 10\% than the one estimated by the \citet{JTP97} method .

The \sbl\ catalogue is too dense for the relative crudeness of the X-ray source positions ($\sigma_{\alpha\delta}=0\farcs7\pm0\farcs3$ on average, $\sigma_{\alpha\delta}$  being defined on 2D). Indeed, even adopting a limited cross-correlation radius of 2\farcs5 would yield over 100 spurious identifications. We thus decided to decrease the limiting magnitude of the catalogue. The maximum of the $\phi_\mathrm{true}$ function is obtained adopting $V<20$. At the distance of the cluster, this corresponds to the magnitude of a M0 dwarf ($M\sim0.5$~\msol). PMS low-mass stars being brighter than ZAMS stars of the same mass, the progenitors of M0 stars should thus still be included in the optical list. Finally, we note that a significant improvement (in terms of the relative percentage of spurious  associations) is obtained when restricting the \sbl\ cross-correlation to objects with $V<19$. The drawback is that the number of true associations is also significantly reduced. Table \ref{tab: jtp97} lists the best fit parameters of Eq.~\ref{eq: jtp97} for both cases and Table \ref{tab: Xcat} provides the \sbl\ cross-identifications down to $V<20$. We leave to the user the choice to restrict the  list to $V<19$ according to his/her motivations.

From the cumulative distributions shown in Fig.~\ref{fig: jtp97}, we adopted the cross-correlation radii corresponding to the knees in the distributions of counterparts;  these are reported in Table \ref{tab: jtp97}.
 The percentage of identified sources ranges from 55 to 83\% according to the catalogue used. The results for the SBL98~v2 and the \sbl\ catalogues are clearly in contrast with the other catalogues. With about 75\% of the total number of X-ray sources in the FOV being identified, among which less than 10\%  statistically correspond to spurious associations, the latter catalogue is probably the most appropriate for the identification processes. 
In the following, we thus adopt the \sbl\ catalogue, that covers the complete \epic\ FOV, as the main reference in the identification of the sources. Table \ref{tab: Xcat_iden} provides the cross identifications between the X-ray source lists and the different optical catalogues. We find at least one counterpart in one catalogue for about 85\% of the X-ray sources. 

While carrying out this work, we noticed some confusion between the names of several sources reported in the widely consulted SIMBAD database. For this reason,  Table \ref{tab: Xcat_iden} also gives other commonly adopted source denominations such as HD, CPD and Braes numbers. The Seggewiss numbering is also extensively used in the literature related to \ngc. We therefore used the original chart of \citet{Se68a} -- subsequently completed by \citet{RCB97} -- and we rederived the cross-correlation to avoid any previous misidentification. 
 
%%%%%%%%%%%%%%%%%%%%%%%%%%%%%%%%%%%%%%%%%%%%%%%%%%%%%%%%%%%%%%%%%%%%%%%%%%%%%%%
%%%%%%%%%%%%%%%%%%%%% IDENTIFICATION TABLE %%%%%%%%%%%%%%%%%%%%%%%%%%%%%%%%%%%%
%%%%%%%%%%%%%%%%%%%%%%%%%%%%%%%%%%%%%%%%%%%%%%%%%%%%%%%%%%%%%%%%%%%%%%%%%%%%%%%

\begin{sidewaystable*}
\begin{minipage}[t][180mm]{\textwidth}
  \caption{Cross-identification of the X-ray source catalogue of Table \ref{tab: Xcat} with the different optical/infrared catalogues: 2MASS, GSC\,2.2, USNO B1.0 and \sbl\ ($V<20$).  $d_\mathrm{cc}$ is the distance (in \arcsec) between the positions of the X-ray source and of its closest counterpart. Only the closest identification has been reported in this table.
  Cross-identification with commonly used denominations is given in the last columns (Col.~26 to 31) of the table.}
\label{tab: Xcat_iden}
\tiny
\begin{tabular}{c c | c c c c c | c c c c c | c c c c c c c}
\hline 
\hline
    &    XMMU J &  \multicolumn{5}{c|}{2MASS}                             &   \multicolumn{5}{c|}{GSC\,2.2}                         &  \multicolumn{7}{c}{USNO B1.0}                                  \\
 X\,\#  & HHMMSS.s$\pm$DDAMAS    &  $d_\mathrm{cc}$  &         name     &    J   &    H   &    K   &  $d_\mathrm{cc}$ &        name    &   R   &   B   &   V   &  $d_\mathrm{cc}$  &     name     &  R1   &  B1   &   R2  &  B2   &   I   \\

 [1] & [2] & [3] & [4] & [5] & [6] & [7] & [8] & [9] & [10] & [11] & [12] & [13] & [14] & [15] & [16] & [17] & [18] & [19] \\
\hline     
  1 & 165300.0$-$415444 &  1.23 & 16530004$-$4154458 & 14.597 & 13.942 & 13.766 & 1.36 & S230322023799  & 16.58 &       & 17.48 & 1.38 & 0480$-$0520459 &       & 15.60 & 17.86 & 16.57 & 14.82        \\         
  2 & 165304.3$-$415334 &       &                    &        &        &        & 2.25 & S230322024061  & 16.90 &       & 17.34 & 1.79 & 0481$-$0502790 &       & 15.92 & 17.73 & 16.91 & 14.95  	    \\
  3 & 165305.2$-$415204 &  1.66 & 16530516$-$4152052 & 15.310 & 14.353 & 13.998 &      &                &       &       &       & 1.32 & 0481$-$0502801 &       & 17.50 & 20.84 & 19.15 &        	    \\
  4 & 165306.9$-$414931 &       &                    &        &        &        &      &                &       &       &       &      &              &       &       &       &       &        	    \\
  5 & 165307.4$-$414659 &       &                    &        &        &        &      &                &       &       &       &      &              &       &       &       &       &        	    \\ 
  6 & 165307.4$-$414345 &  1.31 & 16530729$-$4143454 &  7.933 &  7.926 &  7.894 & 1.32 & S2303220656    &       &  8.71 &  8.50 & 1.33 & 0482$-$0494626 &  8.69 &  8.39 &  8.58 &  8.37 &  8.29  	    \\
  7 & 165308.1$-$414533 &  0.51 & 16530820$-$4145336 & 15.459 & 14.560 & 14.247 &      &                &       &       &       &      &                &       &       &       &       &        	    \\
  8 & 165310.3$-$414733 &  1.28 & 16531017$-$4147340 & 13.148 & 12.553 & 12.271 & 1.13 & S230322025106  & 15.97 &       & 16.68 & 1.14 & 0482$-$0494652 &       & 15.36 & 17.13 & 15.92 & 13.87  	    \\
  9 & 165310.8$-$414451 &  0.99 & 16531068$-$4144512 & 14.126 & 13.467 & 13.234 & 1.28 & S230322025351  & 16.32 &       & 16.96 & 0.80 & 0482$-$0494657 &       & 15.59 & 17.33 & 16.49 & 14.87  	    \\
 10 & 165311.5$-$414755 &  2.94 & 16531161$-$4147582 & 15.438 & 14.490 & 14.349 &      &                &       &       &       &      &                &       &       &       &                 &    \\
 11 & 165312.9$-$415049 &  0.76 & 16531300$-$4150489 & 12.821 & 12.395 & 12.018 & 1.17 & S230322024617  & 14.78 &       & 15.16 & 0.60 & 0481$-$0502869 &       & 14.40 & 15.71 & 14.28 & 13.17  	    \\
 12 & 165313.2$-$415222 &  0.96 & 16531315$-$4152218 & 13.426 & 12.709 & 12.393 & 1.56 & S230322024317  & 15.87 &       & 16.36 & 1.12 & 0481$-$0502872 &       & 15.12 & 16.78 & 15.81 & 14.14  	    \\
 13 & 165313.6$-$415133 &  1.32 & 16531344$-$4151336 & 13.826 & 13.209 & 13.036 & 1.56 & S230322024479  & 15.81 &       & 16.43 & 1.00 & 0481$-$0502879 &       & 15.20 & 16.75 &       & 14.29  	    \\
 14 & 165315.3$-$415011 &  1.03 & 16531528$-$4150113 & 14.925 & 14.070 & 13.681 & 1.25 & S2303220106701 & 17.53 &       &       & 0.88 & 0481$-$0502898 &       & 16.93 &       & 17.92 & 15.75  	    \\
 15 & 165315.8$-$414436 &       &                    &        &        &        & 0.63 & S2303220108158 & 17.51 &       &       & 0.39 & 0482$-$0494707 &       & 16.55 & 19.18 & 17.80 & 15.62  	    \\
 16 & 165316.3$-$415139 &  2.48 & 16531618$-$4151423 & 14.891 & 14.169 & 13.907 & 2.25 & S2303220106219 & 17.61 &       &       & 2.18 & 0481$-$0502907 &       & 16.83 & 19.16 & 17.66 & 15.72  	    \\
 17 & 165316.5$-$415024 &  2.34 & 16531643$-$4150266 & 15.897 & 15.118 & 14.763 &      &                &       &       &       &      &                &       &       &       &       &        	    \\
 18 & 165317.0$-$414412 &  1.30 & 16531694$-$4144114 & 13.172 & 12.505 & 12.272 & 1.56 & S230322025408  & 15.44 &       & 15.83 & 1.31 & 0482$-$0494722 &       & 14.81 & 16.75 & 15.32 & 14.00  	    \\
 19 & 165318.4$-$415215 &  1.29 & 16531851$-$4152161 & 15.297 & 14.335 & 14.006 &      &                &       &       &       & 1.68 & 0481$-$0502931 &       &       &       & 20.38 & 16.41  	    \\
 20 & 165318.7$-$415502 &  0.30 & 16531869$-$4155025 & 15.138 & 14.189 & 13.846 &      &                &       &       &       & 0.12 & 0480$-$0520711 &       & 17.47 & 21.34 & 19.45 & 16.63  	    \\
 21 & 165318.9$-$415446 &       &                    &        &        &        &      &                &       &       &       &      &                &       &       &       &       &        	    \\
 22 & 165319.4$-$415431 &  1.17 & 16531939$-$4154299 &  9.650 &  9.546 &  9.478 & 0.70 & S2300110209    & 10.27 &       & 10.58 & 0.68 & 0480$-$0520725 &       & 10.23 & 11.16 & 10.69 & 10.06  	    \\
 23 & 165320.6$-$414249 &       &                    &        &        &        & 1.48 & S230322025513  & 15.86 &       & 16.49 & 1.00 & 0482$-$0494765 &       & 15.05 & 17.99 & 15.69 & 14.02  	    \\ 
 24 & 165322.0$-$414109 &  0.44 & 16532192$-$4141093 & 14.806 & 13.970 & 13.719 & 0.71 & S2303220108655 & 17.66 &       &       & 0.12 & 0483$-$0487362 &       & 17.04 & 21.08 & 17.94 & 15.36  	    \\
 25 & 165324.2$-$414720 &  1.66 & 16532414$-$4147222 & 11.989 & 11.567 & 11.458 & 1.61 & S230322025129  & 13.49 &       & 13.86 &      &                &       &       &       &       &        	    \\

\hline 
\end{tabular}
\vspace*{5mm}\\
\begin{tabular}{c c | c c c c c c | c c c c c c}
\hline 
\hline
    &    XMMU J   &    \multicolumn{6}{c|}{\sbl\ ($V<20$)} & \multicolumn{6}{c}{Other denominations}\\
 X\,\#  &  HHMMSS.s$\pm$DDAMAS   &  $d_\mathrm{cc}$ & name  &   V    &  V$-$I   &  B$-$V  &  U$-$B & HD/HDE & CD & CPD & Segg. & SBL98 & Braes \\

 [1] & [2]              &   [20] & [21]   &   [22]  &  [23] &  [24] &   [25]  & [26] & [27] & [28] & [29] & [30] & [31] \\
\hline     
  1 & 165300.0$-$415444 &   1.23 &    928 &  17.515 & 1.699 & 1.336 &         &      &      &      &      &      &     \\
  2 & 165304.3$-$415334 &   2.24 &   1065 &  17.960 & 2.047 & 1.541 &         &      &      &      &      &      &     \\ 
  3 & 165305.2$-$415204 &   1.64 &  13015 &  19.757 & 2.755 &       &         &      &      &      &      &      &     \\ 
  4 & 165306.9$-$414931 &        &        &         &       &       &         &      &      &      &      &      &     \\ 
  5 & 165307.4$-$414659 &        &        &         &       &       &         &      &      &      &      &      &     \\ 
  6 & 165307.4$-$414345 &   1.16 &   1147 &   8.471 & 0.358 & 0.240 &$-$0.731 &      &      &      &      &      &     \\ 
  7 & 165308.1$-$414533 &        &        &         &       &       &         &      &      &      &      &      &     \\ 
  8 & 165310.3$-$414733 &   1.16 &   1221 &  17.071 & 2.570 & 1.488 &         &      &      &      &      &      &     \\ 
  9 & 165310.8$-$414451 &   0.92 &   1239 &  17.140 & 1.743 & 1.398 &   0.742 &      &      &      &      &      &     \\ 
 10 & 165311.5$-$414755 &        &        &         &       &       &         &      &      &      &      &      &     \\  
 11 & 165312.9$-$415049 &   0.87 &   1306 &  15.316 & 1.486 & 1.137 &   0.531 &      &      &      &      &      &     \\ 
 12 & 165313.2$-$415222 &   1.56 &   1311 &  16.422 & 1.658 & 1.223 &   0.549 &      &      &      &      &      &     \\
 13 & 165313.6$-$415133 &   1.47 &   1324 &  16.553 & 1.652 & 1.244 &   0.615 &      &      &      &      &      &     \\
 14 & 165315.3$-$415011 &   1.08 &  13784 &  18.713 & 2.276 & 1.596 &         &      &      &      &      &      &     \\
 15 & 165315.8$-$414436 &   0.39 &  13821 &  18.372 & 1.990 & 1.709 &         &      &      &      &      &      &     \\
 16 & 165316.3$-$415139 &   2.36 &  13851 &  18.519 & 2.179 & 1.680 &         &      &      &      &      &      &     \\
 17 & 165316.5$-$415024 &        &        &         &       &       &         &      &      &      &      &      &     \\
 18 & 165317.0$-$414412 &   1.37 &   1426 &  16.153 & 1.698 & 1.370 &   0.753 &      &      &      &      & 3021 &     \\
 19 & 165318.4$-$415215 &        &        &         &       &       &         &      &      &      &      &      &     \\
 20 & 165318.7$-$415502 &   0.19 &  14041 &  19.530 & 2.547 &       &         &      &      &      &      &      &     \\
 21 & 165318.9$-$415446 &        &        &         &       &       &         &      &      &      &      &      &     \\
 22 & 165319.4$-$415431 &   1.37 &   1500 &  10.598 & 0.610 & 0.416 &$-$0.437 &      &      &      &      &   19 &     \\
 23 & 165320.6$-$414249 &   1.10 &   1535 &  16.870 & 1.894 & 1.496 &   0.926 &      &      &      &      & 3075 &     \\
 24 & 165322.0$-$414109 &   0.52 &  14282 &  18.967 & 2.472 & 1.707 &         &      &      &      &      & 3119 &     \\
 25 & 165324.2$-$414720 &   1.40 &   1615 &  13.892 & 1.132 & 0.876 &   0.259 &      &      &      &      &   33 &     \\
\hline
\end{tabular}
\vfill
\end{minipage}
\end{sidewaystable*}
%%%%%%%%%%%%%%%%%%%%%%%%%%%%%%%%%%%%%%%%%%%%%%%%%%%%%%%%%%%%%%%%%%%%%%%%%%%%%%%
%%%%%%%%%%%%%%%%%%%%%%%%%%%%%%%%%%%%%%%%%%%%%%%%%%%%%%%%%%%%%%%%%%%%%%%%%%%%%%%
%%%%%%%%%%%%%%%%%%%%%%%%%%%%%%%%%%%%%%%%%%%%%%%%%%%%%%%%%%%%%%%%%%%%%%%%%%%%%%%

%--------------------------------------------------------------------------------------------------------------------------

\subsection{The detection limit}

This paragraph aims at the evaluation of the detection limit of the present X-ray catalogue. Though essential, this question is far from trivial because the detection limit is, {\it a priori}, not uniform throughout the field of view. Besides the areas where the detectors do not overlap and the presence of gaps between the detector CCDs, the \xmm\ effective exposure duration  is decreasing from the FOV  centre towards its edges. In addition, both the crowdedness of the field in its central part and the numerous bright sources preferentially located in the core of the cluster also affect the detection limit in a non uniform way. As an approximation we neglect the effects of the gaps, mainly focusing on the three other effects.

The exposure maps computed for the three \epic\ instruments and their different combinations display a smooth decrease from the centre of the detector to its edges by about a factor of three. In terms of the amount of signal collected for two similar sources -- one located near the FOV axis, the other near one of its edges -- the number of counts $n$ received will be three times higher near the axis. Neglecting any background effect, the signal-to-noise ratio is approximately  given by $S/N=\sqrt{n}$.  For the outer source, it is therefore smaller by a factor of $\sqrt{3} \approx 1.7$. To the first order, the detection limit in the outer parts of the field is thus about a factor two higher than in the central part of the FOV. As a next step, we used the SAS task {\it esensmap} to build sensitivity maps corresponding to the current exposure maps and to the adopted logarithmic likelihood detection thresholds $L_2$. The sensitivity maps obtained actually provide the minimum number of counts for a source to be detected by the detection task \eml\ according to the given equivalent logarithmic likelihood threshold. These maps indeed  predict that the sensitivity of the \epic\ camera is twice larger near the axis than in the outer parts of the detector whatever the instrument combination is. This is in agreement with our previous estimate.

\begin{figure}
   \centering
   \includegraphics[width=8cm]{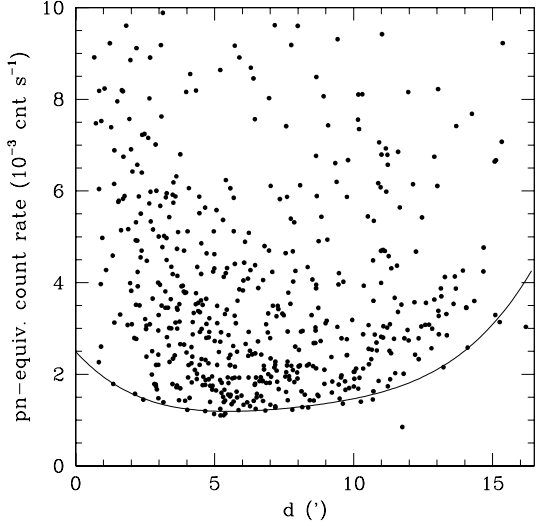}
   \caption{Bottom part of the distribution of the {\it pn-equivalent} count rates of the X-ray sources as a function of their distance ($d$) to \hda. The plain line shows the adopted lower limit given by Eq.~\ref{eq: pnvsd}.}
   \label{fig: pnvsd}
\end{figure}

Accounting for the variation of the source density and the distribution of the bright sources in the FOV is a more tricky issue. We chose to adopt a completely empirical approach, taking advantage of the large number of X-ray sources in the field. We assumed that a very good indication of the detection limit in the different parts of the field is given by the brightness of the faintest sources detected in these selected areas. We adopted the following approach. Because of the presence of gaps, we computed an {\it equivalent} \epicpn\ count rate, in the range 0.5-10.0~keV, for each source. To the first order, the relation between the count rates measured in any of the two \mos\ detectors and in the \pn\ detector is approximately linear. Using the count rates obtained for sources that were detected on several \epic\ instruments, we thus derived an empirical conversion factor between the \mos1, \mos2  and \pn\ count rates. These factors were then used to obtain the so-called {\it pn-equivalent} count rates for all sources and, in particular, for those that fall in the gaps of one or several instruments. This yields approximately coherent count rates for the different sources, whatever their position on the detectors. Figure \ref{fig: pnvsd} displays the source {\it pn-equivalent} count rates as a function of the distance from the central axis of the FOV -- assumed to be the position of the binary \hda. A lower limit is clearly seen in the distribution. Selecting the faintest sources (i.e. those displaying the lowest {\it equivalent} count rates) in successive rings centered on \hda\ provides an approximate sampling of this limit. We then adjusted a polynomial and  derived an empirical detection limit in terms of {\it pn-equivalent} count rates ($cr_\mathrm{lim.}$) as a function of the distance ($d$) from the field axis. This limit (in units of $10^{-3}$\,\cnts) is described by the following relation:

\begin{figure}
   \centering
   \includegraphics[width=8cm]{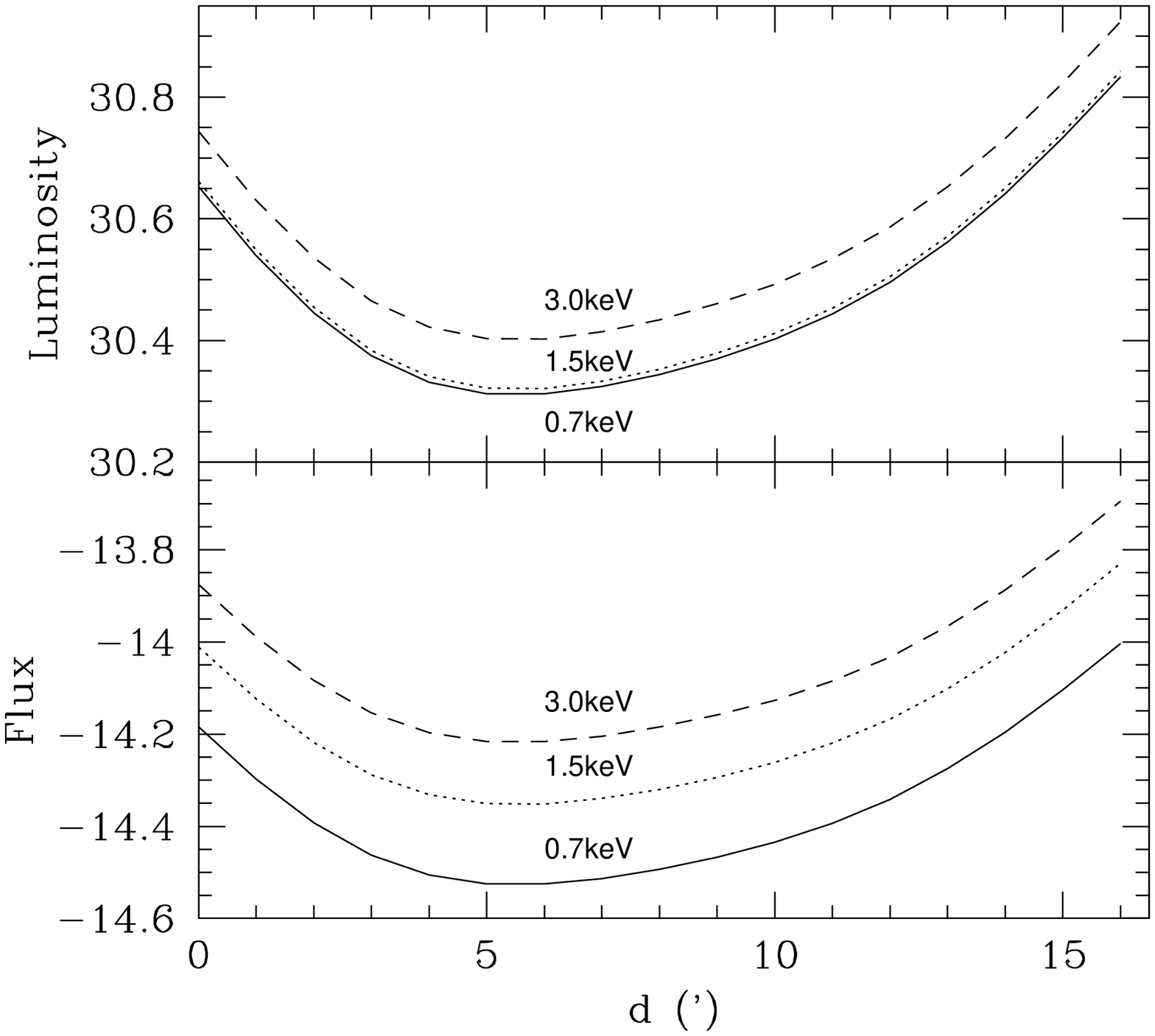}
   \caption{{\bf Lower panel:} Estimated detection limit  expressed in terms of the observed flux (in $\log($\ergscm)). {\bf Upper panel:} Equivalent detection limit, expressed in ISM-absorption corrected luminosity (in $\log$(\ergs)), for sources located in the \ngc\ cluster ($DM=11.07$, $n_\mathrm{H,ISM}=0.26\times10^{22}$~cm$^{-2}$). The different lines refer to the different  \mekal\ model temperatures adopted for the conversion. The energy band considered in both panels is 0.5-10.0\,keV. }
   \label{fig: detlim}
\end{figure}

\begin{equation}
\begin{array}{c c c c c c c}
cr_\mathrm{lim.}(d) 
&  = & 2.49214 & - & 0.65577\ d &  + & 0.11822\ d^2 \\
& - & 0.00929\ d^3 &  + & 0.00030\ d^4
\label{eq: pnvsd}
\end{array}
\end{equation}
where $d$ is the distance to \hda\ expressed in arcmin. Eq.~\ref{eq: pnvsd}  is shown in Fig.~\ref{fig: pnvsd}. Clearly the detection limit is higher in the central part of the field ($d<5\arcmin$), most probably because of the higher source density and because bright sources are preferentially located in the inner part of the FOV. The sensitivity also decreases towards the CCD edges, as indicated both by the exposure maps and the sensitivity maps. Finally we used single temperature optically thin thermal plasma  models of the Raymond-Smith type to convert the {\it pn-equivalent} count rates given by Eq.~\ref{eq: pnvsd} to fluxes and luminosities. For this purpose, we adopted the conversion computed by the WebPIMMS converter\footnote{WebPIMMS is a NASA's HEASARC tool powered by PIMMS v3.6a. It is hosted at the following URL:  http://heasarc.gsfc.nasa.gov/ Tools/w3pimms.html}, assuming a column density of $0.26\times10^{22}$\,cm$^{-2}$, typical of the interstellar absorbing column for the cluster. Results are displayed in Fig.~\ref{fig: detlim} for three different plasma temperatures. In conclusion, the flux detection limit is approximately located between $3\times10^{-15}$ and $1.5\times10^{-14}$~\ergscm, depending on the distance from the detector axis and on the source spectrum. In the central part of the FOV, we consider that the typical limiting flux is about $6\times10^{-15}$~\ergscm\ for soft sources.

\begin{figure}
   \centering 
   \includegraphics[angle=-90,width=8cm]{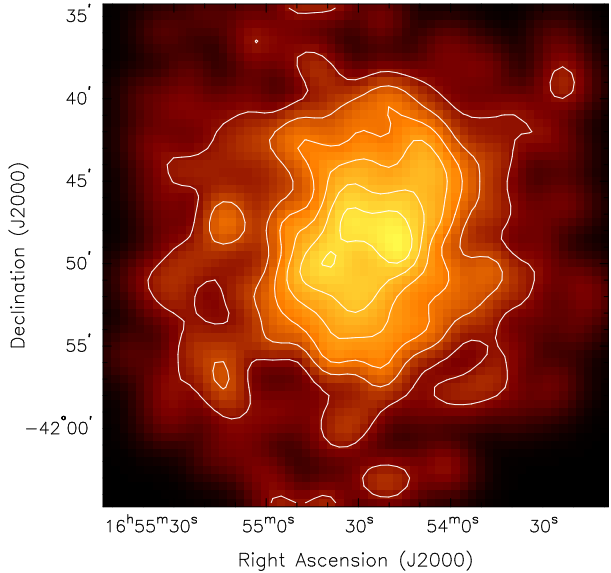}
   \caption{2-D distribution of the surface density of the X-ray sources. The image is centered on the location of \hda. The original source density chart was convolved with a Gaussian with $\sigma=1\arcmin$. Overplotted contour levels correspond to 1, 1.5, 2, 2.5, 3 and 3.5 sources per arcmin$^2$.  }
   \label{fig: 2Ddens}
   \end{figure}

%__________________________________________________________________

\section[]{The X-ray sources in \ngc} \label{sect: Xsources}

This section presents an overview of the main properties of the X-ray sources in \ngc. No attempt will be made here to investigate the characteristics of the different sub-populations of the cluster. This work is postponed to future devoted papers.

\subsection{Spatial distribution }\label{ssect: space}

As seen from Figs. \ref{fig: ngc6231} to \ref{fig: dss_5}, there is an obvious clustering of the X-ray sources in the inner part of the FOV. Their spatial distribution projected on the sky presents, at first sight, a revolution symmetry around the centre of the field, i.e. the position of \hda. Considering the sources located at less than 15\arcmin\ from \hda, we computed the geometrical centre of the source distribution. We also computed the {\it brightness} centre of the X-ray image. For this purpose, we adopted the {\it pn-equivalent} count rates for each source. The two centres are located slightly East of \hda, at no more than 30\arcsec\ (see Fig.~\ref{fig: dss_5}). From the two-dimensional map of the X-ray source density (Fig.~\ref{fig: 2Ddens}), we conclude that there is only a slight deviation from this scheme and that the X-ray source distribution shows a slight N-S elongation. In the following, we however  assume  that the distance from the cluster centre, i.e.\ from \hda, remains the main parameter that shapes the source distribution. We also adopt the position of \hda\ as the  very centre of the cluster.

\begin{figure}
   \centering
   \includegraphics[width=8cm]{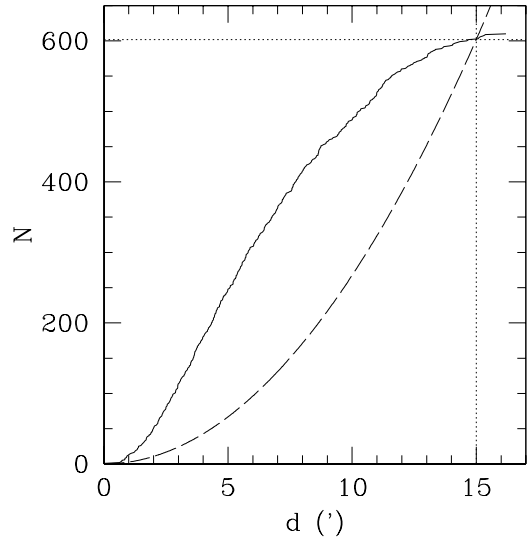}
   \caption{{\bf Plain line:} cumulative distribution of the number of X-ray sources ($N$) with increasing distance ($d$) from \hda. {\bf Dashed line:} idem, but computed assuming a uniform distribution of the 602 sources (dotted horizontal line) located within a 15\arcmin\ (dotted vertical line) circle around \hda. }
   \label{fig: NvsRad}
\end{figure}

\begin{figure}
   \centering
   \includegraphics[width=8cm]{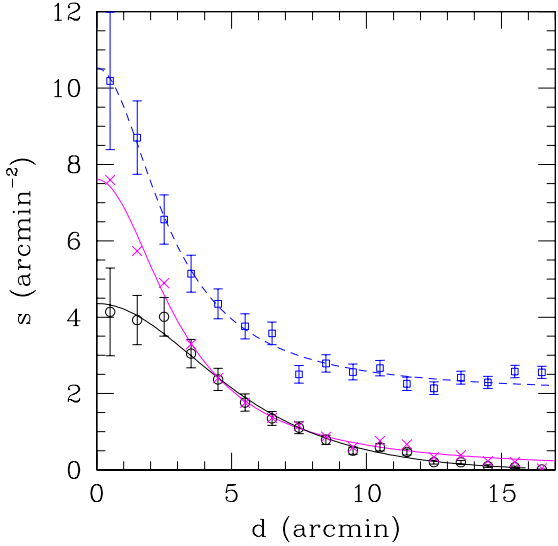}
   \caption{Surface density profiles of the X-ray sources (open circles) and of the optical sources (open squares) with $V<17$. Crosses indicate the X-ray density profile corrected for the empirical \epic\ sensitivity curve (see text). Best-fit King profiles are overplotted. }
   \label{fig: Sdens}
\end{figure}

From Figs. \ref{fig: 2Ddens} and \ref{fig: NvsRad}, it is clear that the radial distribution of the sources is not uniform and that most of them lie within a 10\arcmin\ radius around the cluster centre. We computed the radial density profile of the X-ray emitters and we adjusted an empirical King density profile \citep{Ki62} for a spherically distributed source population :
\begin{equation}
f(d)=k \left[ 1/ \sqrt{1+\left(d/d_\mathrm{c} \right)^2} -  1/ \sqrt{1+\left(d_\mathrm{t}/d_\mathrm{c}\right)^2} \right]^2
\end{equation}
where $k$ is the central density, $d_\mathrm{c}$ the core radius and  $d_\mathrm{t}$ the limiting radius. The King profile is very sensitive to $k$ and $d_\mathrm{c}$, but less sensitive to $d_\mathrm{t}$ which is indeed less meaningful for open clusters in the Galactic plane. The best fit parameters are $k=8.9$\,arcmin$^{-2}$, $d_\mathrm{c}=6\farcm5$ and $d_\mathrm{t}=20\farcm5$. As indicated by Figs. \ref{fig: pnvsd} and \ref{fig: detlim}, our detection limit depends on the location of the source on the detector. In a second step, we thus applied a relative correction to the X-ray density profile, accounting for the sensitivity difference as a function of the distance to the detector axis (crosses in Fig.~\ref{fig: Sdens}). The profile is now sharper and is described by: $k=7.6$\,arcmin$^{-2}$,  $d_\mathrm{c}=3\farcm1$ and $d_\mathrm{t}=1.5\times10^3$~arcmin.
 In Fig.~\ref{fig: Sdens}, we also present the density profile of the stars in \sbl.  Investigating the source density distribution as a function of the limiting magnitude of the catalogue and of the distance to the detector axis, we further note that the \sbl\ catalogue is almost undoubtedly incomplete in the field centre above $V=18$. This is easily explained by the number of bright sources ($V\approx5-10$, see Fig.~\ref{fig: dss_5}) in this region, that renders the detection of faint sources more difficult. For this reason, Fig.~\ref{fig: Sdens} is restricted to objects brighter than 17 in the $V$ band. \ngc\ is further embedded in the \sco\ association. As a consequence, the surface density does not drop to zero in the outer region of the field. We thus subtracted a threshold of $2$\,arcmin$^{-2}$ prior to the adjustment. King best-fit values are this time $k=8.6$\,arcmin$^{-2}$, $d_\mathrm{c}=2\farcm7$ and $d_\mathrm{t}=1.4\times10^3$\,arcmin. From Fig.~\ref{fig: Sdens}, the correlation between the X-ray and optical surface density profile is obvious and yields similar core radii for \ngc. It further suggests that most of the detected X-ray emitters are physically belonging to \ngc. \\

 As discussed in e.g. \citet{SBC04}, X-ray emission is probably one of the best membership criterion for young  stars in open clusters. The present X-ray observations probably provide the best census so far of PMS stars in \ngc; though this census is probably still incomplete. However, the \ngc\ X-ray sample might be contaminated by foreground (field stars)  and background (AGNs) objects.
As a last check, we thus roughly estimated the probable number of foreground and background X-ray sources detected in the present campaign. Starting with the foreground objects, we proceeded as explained below. Accounting for the different typical X-ray luminosities for field stars of spectral type O to M and for our flux detection limit, we estimated the maximum distance at which a star can be located while still being detected. Using the so-derived distance, we computed the volume projected onto the \xmm\ FOV. As a last step, we adopted typical star densities in the solar neighbourhood as  quoted by \citet{All73} for the different spectral types. We finally end up with about 20 foreground X-ray sources, most of which are expected to be G-type objects (12 stars) and F-type dwarves (4 or 5 stars). However, the previous approach does not account for probable active stars or RS~CVn in the FOV, which have lower spatial densities but much higher luminosities. Using the work of \citet{Mak03}, we found that about 21 galactic active stars could be detected in the \epic\ FOV. This yields a total of approximatively 41 contaminating galactic sources. As an additional check, we also used the X-ray stellar $\log N(>S)-\log S$ curve at low galactic latitudes provided by \citet{MHG03}. Again we found that about 40 galactic X-ray sources are to be expected within our \epic\ FOV.

\begin{figure}
   \centering
   \includegraphics[width=8cm]{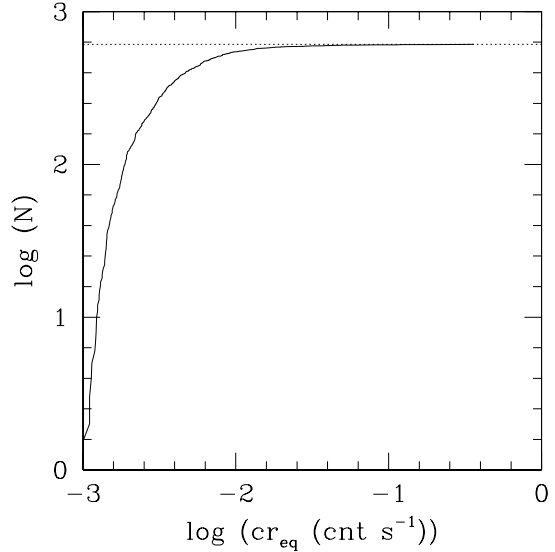}
   \caption{Cumulative distribution of the number of sources as a function of increasing {\it pn-equivalent} count rate. The horizontal dotted line indicates the total number of sources in the FOV.}
   \label{fig: NvsCR}
\end{figure}

We then obtained a rough estimate of the number of extragalactic background sources in our EPIC field. The Galactic coordinates of the cluster are $l_{\rm II} = 343\fdg46$, $b_{\rm II} = +1\fdg19$. Therefore, the total neutral hydrogen column density along this direction must be extremely large and should produce a substantial absorption of X-ray photons from extragalactic sources. Although they are in principle not suited for lines of sight at $|b_{\rm II}| \leq 5^{\circ}$, we used the {\it DIRBE/IRAS} extinction maps provided by  \citet{SFD98} to derive a first order estimate of the total column density. In this way, we find a total Galactic $E(B-V)$ of about 5.6\,mag. Using the gas to dust ratio of \citet{BSD78}, we thus estimate a neutral hydrogen column density of $\sim 3.2 \times 10^{22}$\,cm$^{-2}$. Assuming that extragalactic background sources have a power-law spectrum with a photon index of 1.4, and suffer a total interstellar absorption of $3.2 \times 10^{22}$\,cm$^{-2}$, the mean detection limit $1.9 \times 10^{-3}$\,cnt\,s$^{-1}$ with the pn camera translates into unabsorbed fluxes of $1.2 \times 10^{-14}$ and $3.5 \times 10^{-13}$\,erg\,cm$^{-2}$\,s$^{-1}$ in the 0.5 -- 2.0\,keV and 2.0 -- 10\,keV band respectively. Using the $\log{N}$ -- $\log{S}$ relation from \citet{GRT01}, one expects thus about 13 -- 16 extragalactic objects among the detected sources. Thus, about 2\% of the total number of sources could be background AGNs. It should be emphasized that these background AGNs are expected to appear as rather hard (i.e.\ heavily absorbed) X-ray sources. 

In summary, both the geometrical and X-ray brightness centres of the set of detected sources correspond to the optical cluster centre. The radial profile of the X-ray source density is well  correlated with the optical source radial profile. Both indicate a cluster core radius close to 3\arcmin. Finally, we expect that less than 10\% of the presently detected sources correspond to foreground or background objects. We thus propose that the large majority of the X-ray emitters revealed by the present \xmm\ campaign are mostly belonging to \ngc. Some of them might alternatively  belong to the \sco\ association, in which \ngc\ is embedded.

\begin{figure*}
   \centering
   \includegraphics[width=5.7cm]{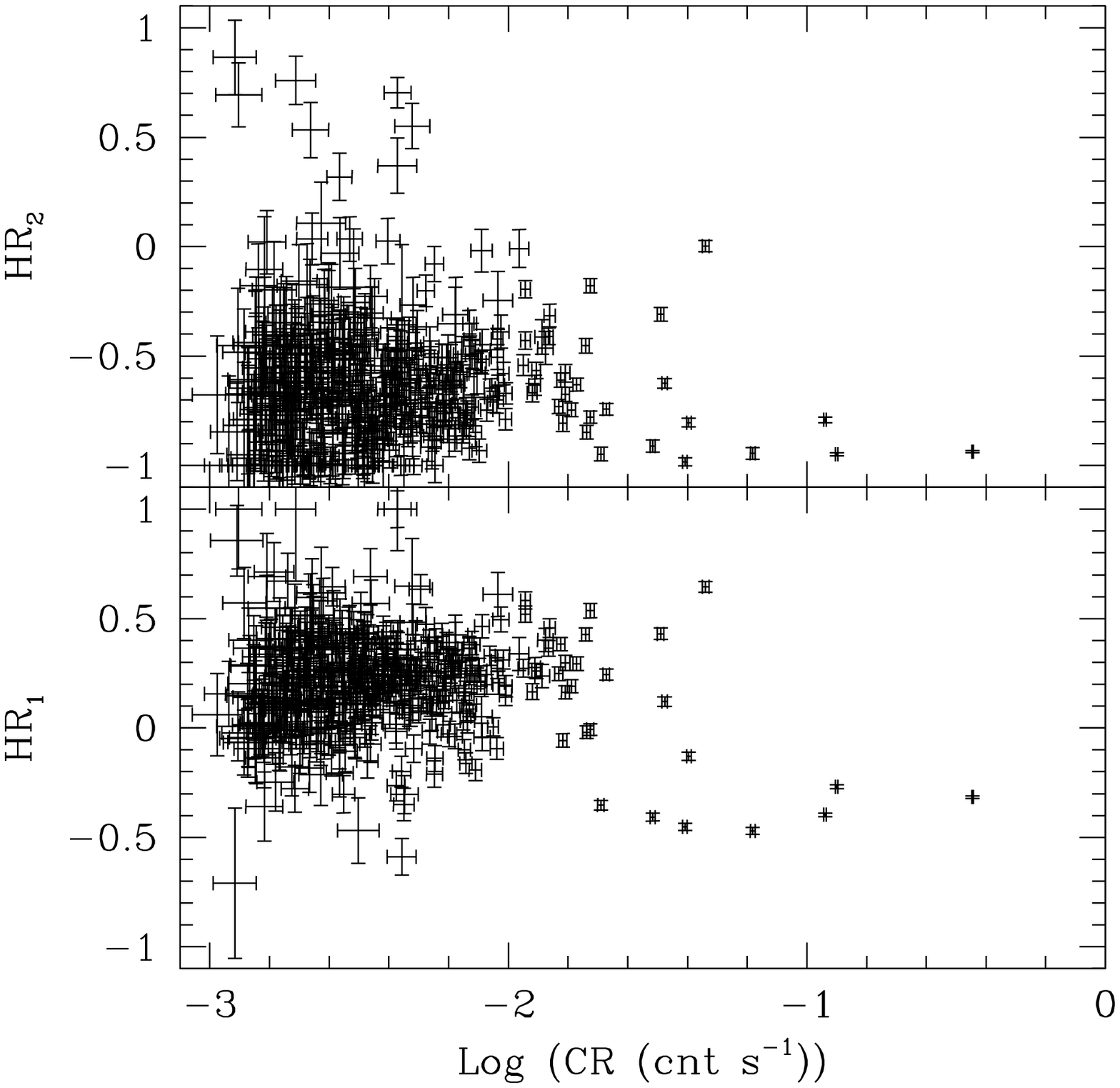}
   \includegraphics[width=5.7cm]{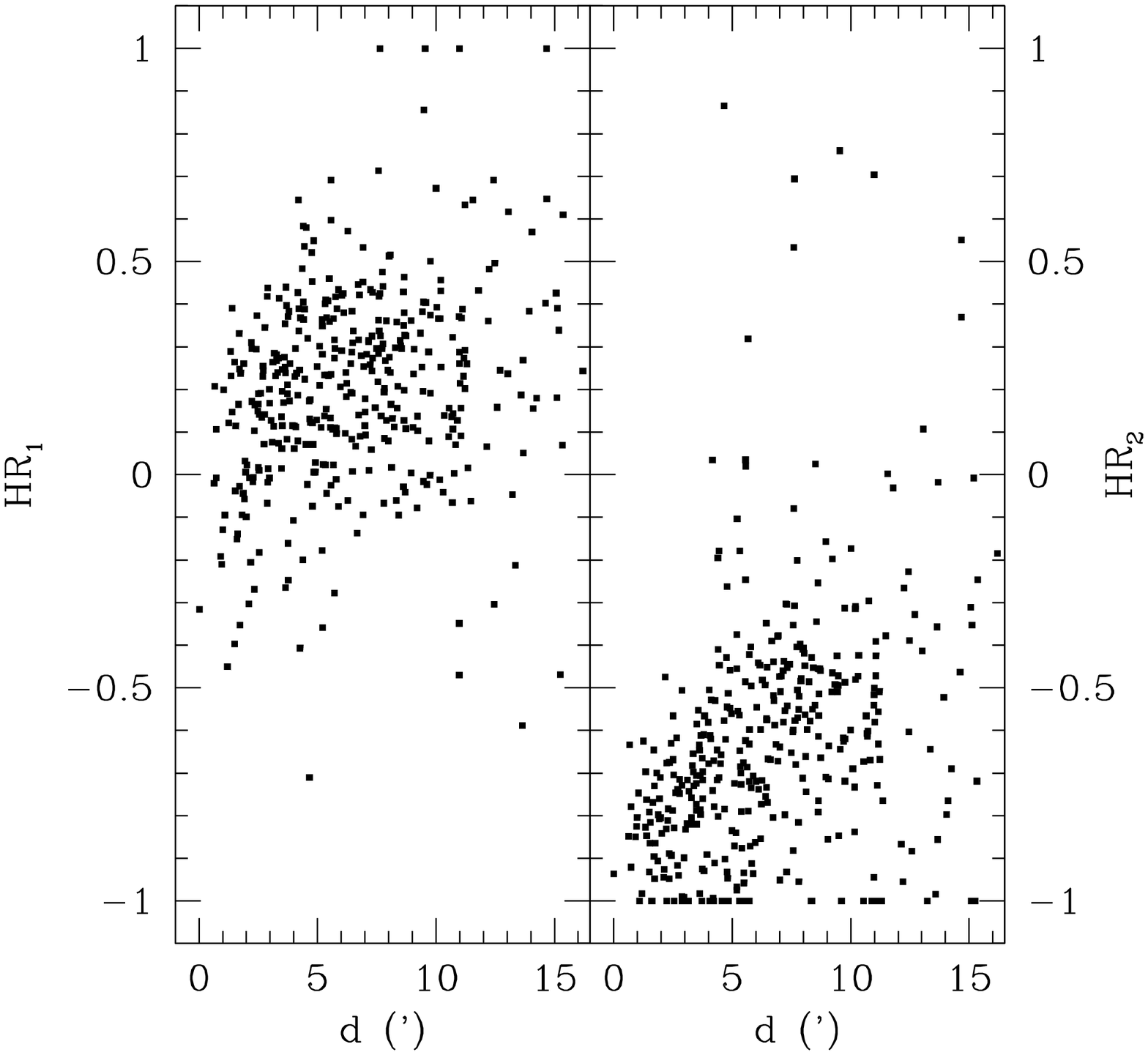}
   \includegraphics[width=5.7cm]{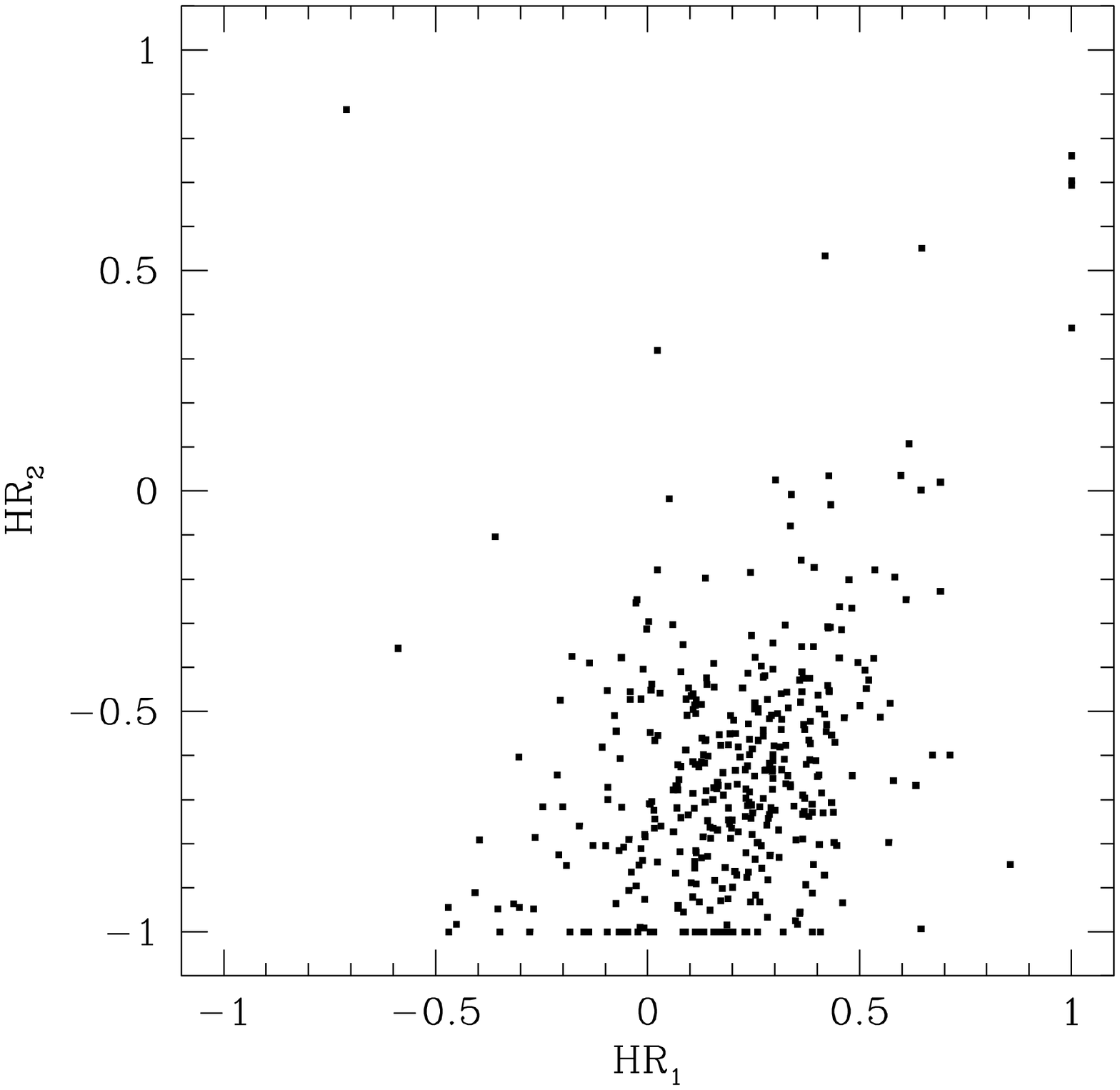}
   \caption{{\bf Left panel:} Hardness ratios versus count rate. {\bf Middle panel:} Hardness ratios versus distance ($d$) from the FOV centre. {\bf Right panel:} $HR_2$ versus $HR_1$. The three panels are built for the sources detected with the \pn\ instrument. Results for the \epicmos\ instruments are similar.}
   \label{fig: HRvs}
\end{figure*}

\subsection{Emission distribution}

\begin{figure}
   \centering
   \includegraphics[width=8cm]{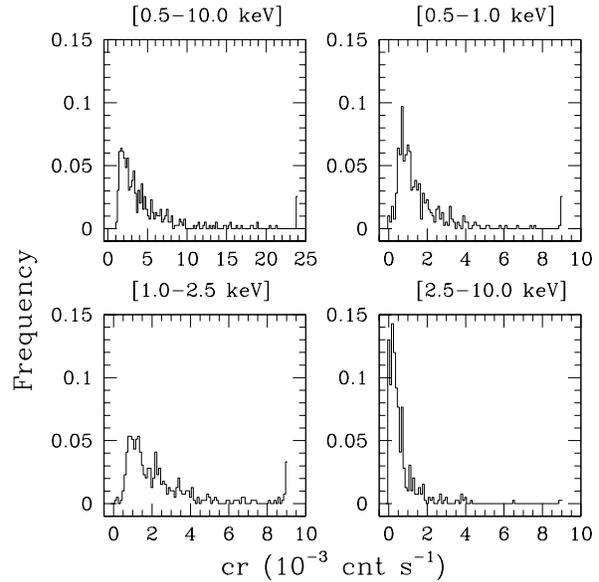}
   \caption{Distribution of the X-ray emitter count rates in the different energy bands considered. The four panels were plotted only using the 392 sources detected with the \pn\ instrument. Results for the \epicmos\ instruments are similar. The last bin includes the contributions of all the brightest X-ray sources.}
   \label{fig: histopn}
\end{figure}

While \hda, the brightest X-ray emitter in the FOV, displays a {\it pn-equivalent} count rate larger than 0.36~\cnts, most of the other sources are much fainter with a count rate below $10^{-2}$\,\cnts (Fig.~\ref{fig: NvsCR}). It is clear from Fig.~\ref{fig: ngc6231} that most of the brightest sources -- associated  with the O-type objects of the cluster -- are relatively soft while the majority of the X-ray emitters have their maximum of emission in the medium band.  Except for the brightest sources, there is no obvious correlation between the source intensity and the source hardness ratios. On average, the detected sources are moderately hard with  $HR_1>0$ and $HR_2<0$ (Fig.~\ref{fig: HRvs}).
The hardness ratios might however show a slight increase towards the edge of the detectors, probably due to the relative dominance of low-mass stars in the outer regions of the FOV.
The histograms of the detected source count rates in the S$_\mathrm{X}$, M$_\mathrm{X}$ and H$_\mathrm{X}$ bands  (Fig.~\ref{fig: histopn}) reveal clear peaks around 0.7, 1.0 and $0.2\times10^{-3}$~\cnts\ respectively. The count rate in the 0.5-10.0~keV band clusters at $2\times10^{-3}$~\cnts\ and the two hardness ratios around 0.2 and $-$0.6 respectively. Accounting for the cluster typical ISM absorbing column $n_\mathrm{H, ISM}=0.26\times10^{22}$~cm$^{-2}$, these values are roughly described by a \mekal\ model with a temperature of k$T=1.0-2.0$\,keV. The corresponding observed flux is about $5\times10^{-15}$~\ergscm. Adopting a distance modulus $DM=11.07$, this yields a luminosity $\log(L_\mathrm{X})\sim30.5$ (\ergs) for a typical X-ray emitter in the cluster.

%__________________________________________________________________

\section{Summary} \label{sect: ccl}
We presented the first results of an \xmm\ campaign on the young open cluster \ngc\ in the \sco\ association.  With an effective cumulated exposure time of  175\,ks in the two \epicmos\ instruments and of 147.5\,ks in the \epicpn, the campaign was split into six successive observations acquired within 5 days. The combined image, built from the data collected by the three \epic\ instruments during the six pointings, reveals an extremely crowded field. Running the SAS task \eml, we detect 610 X-ray sources among which 19 are reported as extended. The latter are probably constituted by non-resolved point-like sources rather than by physically extended sources. We present  an X-ray catalogue covering the \xmm\ FOV and we cross-correlate it with several optical/infrared catalogues. We find at least one optical and/or infrared counterpart for more than 85\% of the X-ray sources within a limited  cross-correlation radius of 3\arcsec\ at maximum. We estimate our detection flux limit to lie between about $3\times10^{-15}$ and $1.5\times10^{-14}$\,\ergscm\ depending on the position on the detectors and on the source spectrum. 

The surface density distribution of the X-ray sources peaks at the centre of the cluster, which we find to be located very near \hda, and presents a slight N-S elongation. Concerning the radial profile of the surface density distribution, over 50\% of the sources are confined within a 6\arcmin\ radius from the cluster centre and about 80\%  within 10\arcmin. The estimated contamination by foreground and background objects is about 9\%. There is a good similarity between this radial profile and the distribution of stars brighter than $V=17$, suggesting that most of the sources physically belong to \ngc. The radial surface density profile of the X-ray sources is well described by a King profile  with a core radius of about  3\arcmin, similar to the one indicated by the  $V<17$ optical source density profile. 

Finally, beside a few bright and soft objects correlated with the O-type stars of the cluster, the large majority of the X-ray population is relatively faint ({\it pn-equivalent} count rate below $10^{-2}$\,\cnts) and displays an intermediate spectrum of a typical temperature probably around 1.0-2.0\,keV. Typical count rates for the sources are 2.0, 0.7, 1.0 and $0.2\times10^{-3}$~\cnts\ respectively in the total energy  band (0.5-10.0~keV), and in the three energy sub-ranges 0.5-1.0, 1.0-2.5 and 2.5-10.0~keV. At the \ngc\ cluster distance, these values roughly correspond to an X-ray luminosity of about $\log(L_\mathrm{X})\sim30.5$ (\ergs).

More detailed investigations of the X-ray properties of the different sub-populations (early-type stars, PMS objects) of the cluster will be presented in subsequent papers in this series. Finally, the X-ray data related to specific early-type binary systems of particular interest are (will be) presented in dedicated papers (see e.g.\ \citealt{SSG04,SAR05,SGR06_219} as well as \citealt{phd}), together with the derivation of their orbital and physical parameters obtained on the basis of an extensive spectral monitoring campaign in the optical domain.

\begin{acknowledgements}
It is a pleasure to thank Dr. W. Seggewiss for providing a copy of his original paper on the cluster photometry \citep{Se68a}, our system manager, A. Detal, for essential assistance in installing and handling the successive SAS versions and Dr. J.-F. Claeskens for sharing numerical routines. We are also grateful to the referee, Dr.\ R.\ Jeffries, for comments and suggestions that helped to improve the overall quality of the paper.
This research is supported in part by contract P5/36 ``P\^ole d'Attraction Interuniversitaire'' (Belgian Federal Science Policy Office) and through the PRODEX XMM and INTEGRAL contracts. 
H. Sung acknowledges the support of the Korea Science and Engineering
Foundation (KOSEF) to the Astrophysical Research Center for
the Structure and Evolution of the Cosmos (ARCSEC$''$) at Sejong University. 

This publication makes use of data products from USNO B1.0, the Two Micron All Sky Survey (2MASS) and the Guide Star catalogue (GSC) II.
 The 2MASS is a joint project of the University of Massachusetts and the Infrared Processing and Analysis Center/California Institute of Technology, funded by the NASA and the NSF.
The GSC~II is a joint project of the Space Telescope Science Institute and the Osservatorio Astronomico di Torino. Additional support is provided by the ESO, the Space Telescope European Coordinating Facility, the International GEMINI project and the ESA Astrophysics Division.
The SIMBAD database and the Vizier catalogue access tool (CDS, Strasbourg, France) have been consulted for the bibliography and for the purpose of object cross-identification. 
\end{acknowledgements}

\bibliographystyle{aa}

\bibliography{../XMM_CAT_PAPER/ngc6231_Xcat}

\begin{thebibliography}{90}
\expandafter\ifx\csname natexlab\endcsname\relax\def\natexlab#1{#1}\fi

\bibitem[{{Allen}(1973)}]{All73}
{Allen}, C.~W. 1973, {Astrophysical quantities} (London: University of London,
  Athlone Press, 3rd ed.)

\bibitem[{{Arenou} \& {Luri}(1999)}]{AL99}
{Arenou}, F., \& {Luri}, X. 1999, in ASP Conf. Ser., Vol. 167, Harmonizing
  Cosmic Distance Scales in a Post-HIPPARCOS Era, ed. D.~{Egret} \& A.~{Heck},
  13

\bibitem[{{Arentoft} {et~al.}(2001){Arentoft}, {Sterken}, {Knudsen},
  {Freyhammer}, {Duerbeck}, {Pompei}, {Delahodde}, \& {Clasen}}]{ASK01}
{Arentoft}, T., {Sterken}, C., {Knudsen}, M.~R., {et~al.} 2001, \aap, 380, 599

\bibitem[{{Balona}(1983)}]{Bal83}
{Balona}, L.~A. 1983, \mnras, 203, 1041

\bibitem[{{Balona}(1992)}]{Bal92}
{Balona}, L.~A. 1992, \mnras, 254, 404

\bibitem[{{Balona} \& {Engelbrecht}(1985)}]{BaE85}
{Balona}, L.~A., \& {Engelbrecht}, C.~A. 1985, \mnras, 212, 889

\bibitem[{{Balona} \& {Laney}(1995)}]{BL95}
{Balona}, L.~A., \& {Laney}, C.~D. 1995, \mnras, 276, 627

\bibitem[{{Baume} {et~al.}(1999){Baume}, {V{\' a}zquez}, \&
  {Feinstein}}]{BVF99}
{Baume}, G., {V{\' a}zquez}, R.~A., \& {Feinstein}, A. 1999, \aaps, 137, 233

\bibitem[{{Bohlin} {et~al.}(1978){Bohlin}, {Savage}, \& {Drake}}]{BSD78}
{Bohlin}, R.~C., {Savage}, B.~D., \& {Drake}, J.~F. 1978, \apj, 224, 132

\bibitem[{{Bok} {et~al.}(1966){Bok}, {Bok}, \& {Graham}}]{BBG66}
{Bok}, B.~J., {Bok}, P.~F., \& {Graham}, J.~A. 1966, \mnras, 131, 247

\bibitem[{{Braes}(1967)}]{Br67}
{Braes}, L.~L.~E. 1967, Bulletin of the Astronomical Institute of the
  Netherlands Supplement Series, 2, 1

\bibitem[{{Breckinridge} \& {Kron}(1963)}]{BK63}
{Breckinridge}, J.~B., \& {Kron}, G.~E. 1963, \pasp, 75, 248

\bibitem[{{Brownlee} \& {Cox}(1953)}]{BC53}
{Brownlee}, R.~R., \& {Cox}, A.~N. 1953, \apj, 118, 165

\bibitem[{{Cash}(1979)}]{Ca79}
{Cash}, W. 1979, \apj, 228, 939

\bibitem[{{Conti} \& {Alschuler}(1971)}]{CA71}
{Conti}, P.~S., \& {Alschuler}, W.~R. 1971, \apj, 170, 325

\bibitem[{{Corcoran}(1996)}]{Cor96}
{Corcoran}, M.~F. 1996, in Revista Mexicana de Astronomia y Astrofisica
  Conference Series, Vol.~5, 54

\bibitem[{{Corcoran}(1999)}]{Cor99}
{Corcoran}, M.~F. 1999, in Revista Mexicana de Astronomia y Astrofisica
  Conference Series, Vol.~8, 131

\bibitem[{{Crawford}(2001)}]{Cr01}
{Crawford}, I.~A. 2001, \mnras, 328, 1115

\bibitem[{{Crawford} {et~al.}(1971){Crawford}, {Barnes}, {Hill}, \&
  {Perry}}]{CBHP71}
{Crawford}, D.~L., {Barnes}, J.~V., {Hill}, G., \& {Perry}, C.~L. 1971, \aj,
  76, 1048

\bibitem[{{Cutri} {et~al.}(2003){Cutri}, {Skrutskie}, {van Dyk}, {Beichman},
  {Carpenter}, {Chester}, {Cambresy}, {Evans}, \& {et al.}}]{2MASS_mnras}
{Cutri}, R.~M., {Skrutskie}, M.~F., {van Dyk}, S., {et~al.} 2003, University of
  Massachusetts and Infrared Processing and Analysis Center (IPAC/California
  Institute of Technology)

\bibitem[{{Damiani} {et~al.}(2004){Damiani}, {Flaccomio}, {Micela},
  {Sciortino}, {Harnden}, \& {Murray}}]{DFM04}
{Damiani}, F., {Flaccomio}, E., {Micela}, G., {et~al.} 2004, \apj, 608, 781

\bibitem[{{den Herder} {et~al.}(2001){den Herder}, {Brinkman}, {Kahn},
  {Branduardi-Raymont}, {Thomsen}, {Aarts}, {Audard}, {Bixler}, \& {at
  al.}}]{denHerder01_rgs_mnras}
{den Herder}, J.~W., {Brinkman}, A.~C., {Kahn}, S.~M., {et~al.} 2001, \aap,
  365, L7

\bibitem[{{Feast} {et~al.}(1961){Feast}, {Stoy}, {Thackeray}, \&
  {Wesselink}}]{FSTW61}
{Feast}, M.~W., {Stoy}, R.~H., {Thackeray}, A.~D., \& {Wesselink}, A.~J. 1961,
  \mnras, 122, 239

\bibitem[{{Feinstein} \& {Ferrer}(1968)}]{FF68}
{Feinstein}, A., \& {Ferrer}, O.~E. 1968, \pasp, 80, 410

\bibitem[{{Feinstein} {et~al.}(2003){Feinstein}, {Mart{\'{\i}}nez}, {Vergne},
  {Baume}, \& {V{\' a}zquez}}]{FMV03}
{Feinstein}, C., {Mart{\'{\i}}nez}, R., {Vergne}, M.~M., {Baume}, G., \& {V{\'
  a}zquez}, R. 2003, \apj, 598, 349

\bibitem[{{Garc{\'{\i}}a} \& {Mermilliod}(2001)}]{GM01}
{Garc{\'{\i}}a}, B., \& {Mermilliod}, J.~C. 2001, \aap, 368, 122

\bibitem[{{Garrison} \& {Schild}(1979)}]{GaS79}
{Garrison}, R.~F., \& {Schild}, R.~E. 1979, \aj, 84, 1020

\bibitem[{{Giacconi} {et~al.}(2001){Giacconi}, {Rosati}, {Tozzi}, {Nonino},
  {Hasinger}, {Norman}, {Bergeron}, {Borgani}, {Gilli}, {Gilmozzi}, \&
  {Zheng}}]{GRT01}
{Giacconi}, R., {Rosati}, P., {Tozzi}, P., {et~al.} 2001, \apj, 551, 624

\bibitem[{{GSC\,2.2}(2001)}]{GSC22}
{GSC\,2.2}. 2001, Space Telescope Science Institute (STScI) and Osservatorio
  Astronomico di Torino

\bibitem[{{Heske} \& {Wendker}(1984)}]{HW84}
{Heske}, A., \& {Wendker}, H.~J. 1984, \aaps, 57, 205

\bibitem[{{Hill} {et~al.}(1974){Hill}, {Crawford}, \& {Barnes}}]{HCB74}
{Hill}, G., {Crawford}, D.~L., \& {Barnes}, J.~V. 1974, \aj, 79, 1271

\bibitem[{{Houck}(1956)}]{Ho56}
{Houck}, T.~E. 1956, Ph.D.~Thesis

\bibitem[{{Jansen} {et~al.}(2001){Jansen}, {Lumb}, {Altieri}, {Clavel}, {Ehle},
  {Erd}, {Gabriel}, {Guainazzi}, {Gondoin}, {Much}, {Munoz}, {Santos},
  {Schartel}, {Texier}, \& {Vacanti}}]{Jansen01_xmm}
{Jansen}, F., {Lumb}, D., {Altieri}, B., {et~al.} 2001, \aap, 365, L1

\bibitem[{{Jeffries} {et~al.}(1997){Jeffries}, {Thurston}, \& {Pye}}]{JTP97}
{Jeffries}, R.~D., {Thurston}, M.~R., \& {Pye}, J.~P. 1997, \mnras, 287, 350

\bibitem[{{King}(1962)}]{Ki62}
{King}, I. 1962, \aj, 67, 471

\bibitem[{{Laval}(1972)}]{Lav72}
{Laval}, A. 1972, \aap, 21, 271

\bibitem[{{Levato} \& {Malaroda}(1980)}]{LM80}
{Levato}, H., \& {Malaroda}, S. 1980, \pasp, 92, 323

\bibitem[{{Levato} \& {Morrell}(1983)}]{LM83}
{Levato}, H., \& {Morrell}, N. 1983, \aplett, 23, 183

\bibitem[{{Levato} {et~al.}(1988){Levato}, {Morrell}, {Garcia}, \&
  {Malaroda}}]{LMGM88}
{Levato}, H., {Morrell}, N., {Garcia}, B., \& {Malaroda}, S. 1988, \apjs, 68,
  319

\bibitem[{{L\"uhrs}(1997)}]{Lu97}
{L\"uhrs}, S. 1997, \pasp, 109, 504

\bibitem[{{Makarov}(2003{\natexlab{a}})}]{Ma03}
{Makarov}, V.~V. 2003{\natexlab{a}}, \aj, 126, 2408

\bibitem[{{Makarov}(2003{\natexlab{b}})}]{Mak03}
{Makarov}, V.~V. 2003{\natexlab{b}}, \aj, 126, 1996

\bibitem[{{Marggraf} {et~al.}(2004){Marggraf}, {Bluhm}, \& {de Boer}}]{MBdB04}
{Marggraf}, O., {Bluhm}, H., \& {de Boer}, K.~S. 2004, \aap, 416, 251

\bibitem[{{Mason} {et~al.}(1998){Mason}, {Gies}, {Hartkopf}, {Bagnuolo},
  {Brummelaar}, \& {McAlister}}]{MGH98}
{Mason}, B.~D., {Gies}, D.~R., {Hartkopf}, W.~I., {et~al.} 1998, \aj, 115, 821

\bibitem[{{Mason} {et~al.}(2001){Mason}, {Breeveld}, {Much}, {Carter},
  {Cordova}, {Cropper}, {Fordham}, {Huckle}, \& {et al.}}]{Mason01_om_mnras}
{Mason}, K.~O., {Breeveld}, A., {Much}, R., {et~al.} 2001, \aap, 365, L36

\bibitem[{{Mathys}(1988)}]{Mat88}
{Mathys}, G. 1988, \aaps, 76, 427

\bibitem[{{Mathys}(1989)}]{Mat89}
{Mathys}, G. 1989, \aaps, 81, 237

\bibitem[{{Mermilliod}(1981)}]{Me81}
{Mermilliod}, J.~C. 1981, \aap, 97, 235

\bibitem[{{Meynet} {et~al.}(1993){Meynet}, {Mermilliod}, \& {Maeder}}]{MMM93}
{Meynet}, G., {Mermilliod}, J.-C., \& {Maeder}, A. 1993, \aaps, 98, 477

\bibitem[{{Monet} {et~al.}(2003){Monet}, {Levine}, {Canzian}, {Ables}, {Bird},
  {Dahn}, {Guetter}, {Harris}, \& {et al.}}]{USNOB10_mnras}
{Monet}, D.~G., {Levine}, S.~E., {Canzian}, B., {et~al.} 2003, \aj, 125, 984

\bibitem[{{Morgan} {et~al.}(1953{\natexlab{a}}){Morgan}, {Gonz{\' a}lez}, \&
  {Gonz{\' a}lez}}]{MGG53}
{Morgan}, W.~W., {Gonz{\' a}lez}, G., \& {Gonz{\' a}lez}, G.
  1953{\natexlab{a}}, \apj, 118, 323

\bibitem[{{Morgan} {et~al.}(1953{\natexlab{b}}){Morgan}, {Whitford}, \&
  {Code}}]{MWC53}
{Morgan}, W.~W., {Whitford}, A.~E., \& {Code}, A.~D. 1953{\natexlab{b}}, \apj,
  118, 318

\bibitem[{{Motch} {et~al.}(2003){Motch}, {Herent}, \& {Guillout}}]{MHG03}
{Motch}, C., {Herent}, O., \& {Guillout}, P. 2003, Astronomische Nachrichten,
  324, 61

\bibitem[{{Penny} {et~al.}(1994){Penny}, {Bagnuolo}, \& {Gies}}]{PBG94}
{Penny}, L.~J., {Bagnuolo}, W.~G., \& {Gies}, D.~R. 1994, Space Science
  Reviews, 66, 323

\bibitem[{{Penny} {et~al.}(1999){Penny}, {Gies}, \& {Bagnuolo}}]{PGB99}
{Penny}, L.~R., {Gies}, D.~R., \& {Bagnuolo}, W.~G. 1999, \apj, 518, 450

\bibitem[{{Perry} {et~al.}(1990){Perry}, {Hill}, {Younger}, \&
  {Barnes}}]{PHYB90}
{Perry}, C.~L., {Hill}, G., {Younger}, P.~F., \& {Barnes}, J.~V. 1990, \aaps,
  86, 415

\bibitem[{{Perry} {et~al.}(1991){Perry}, {Hill}, \& {Christodoulou}}]{PHC91}
{Perry}, C.~L., {Hill}, G., \& {Christodoulou}, D.~M. 1991, \aaps, 90, 195

\bibitem[{{Preibisch} \& {Zinnecker}(2004)}]{PZ04}
{Preibisch}, T., \& {Zinnecker}, H. 2004, \aap, 422, 1001

\bibitem[{{Raboud}(1996)}]{Rab96}
{Raboud}, D. 1996, \aap, 315, 384

\bibitem[{{Raboud} \& {Mermilliod}(1998)}]{RM98}
{Raboud}, D., \& {Mermilliod}, J.-C. 1998, \aap, 333, 897

\bibitem[{{Raboud} {et~al.}(1997){Raboud}, {Cramer}, \& {Bernasconi}}]{RCB97}
{Raboud}, D., {Cramer}, N., \& {Bernasconi}, P.~A. 1997, \aap, 325, 167

\bibitem[{{Rauw} {et~al.}(2002){Rauw}, {Naz{\' e}}, {Gosset}, {Stevens},
  {Blomme}, {Corcoran}, {Pittard}, \& {Runacres}}]{RNG02}
{Rauw}, G., {Naz{\' e}}, Y., {Gosset}, E., {et~al.} 2002, \aap, 395, 499

\bibitem[{{Rauw} {et~al.}(2003){Rauw}, {De Becker}, {Gosset}, {Pittard}, \&
  {Stevens}}]{RdBG03}
{Rauw}, G., {De Becker}, M., {Gosset}, E., {Pittard}, J.~M., \& {Stevens},
  I.~R. 2003, \aap, 407, 925

\bibitem[{{Sana}(2005)}]{phd}
{Sana}, H. 2005, PhD thesis, Li\`ege University

\bibitem[{{Sana} {et~al.}(2001){Sana}, {Rauw}, \& {Gosset}}]{SRG01}
{Sana}, H., {Rauw}, G., \& {Gosset}, E. 2001, \aap, 370, 121

\bibitem[{{Sana} {et~al.}(2002){Sana}, {Rauw}, {Gosset}, \& {Vreux}}]{SRG02}
{Sana}, H., {Rauw}, G., {Gosset}, E., \& {Vreux}, J.-M. 2002, in ASP Conf.
  Ser., Vol. 260, Interacting Winds from Massive Stars, ed. A.~{Moffat} \&
  N.~{St-Louis}, 431
\bibitem[{{Sana} {et~al.}(2003){Sana}, {Hensberge}, {Rauw}, \&
  {Gosset}}]{SHRG03}
{Sana}, H., {Hensberge}, H., {Rauw}, G., \& {Gosset}, E. 2003, \aap, 405, 1063

\bibitem[{{Sana} {et~al.}(2004){Sana}, {Stevens}, {Gosset}, {Rauw}, \&
  {Vreux}}]{SSG04}
{Sana}, H., {Stevens}, I.~R., {Gosset}, E., {Rauw}, G., \& {Vreux}, J.-M. 2004,
  \mnras, 350, 809

\bibitem[{{Sana} {et~al.}(2005{\natexlab{a}}){Sana}, {Antokhina}, {Royer},
  {Manfroid}, {Gosset}, {Rauw}, \& {Vreux}}]{SAR05}
{Sana}, H., {Antokhina}, E., {Royer}, P., {et~al.} 2005{\natexlab{a}}, \aap,
  441, 213

\bibitem[{{Sana} {et~al.}(2005{\natexlab{b}}){Sana} {Rauw}, \& {Gosset}}]{SRG05}
{Sana}, H., {Rauw}, G. \& {Gosset}, E.  2005{\natexlab{b}}, in {Massive Stars and High-Energy Emission in OB Associations},  Proc.\ JENAM 2005 Distant Worlds, ed. G.~{Rauw} et al., 107

\bibitem[{{Sana} {et~al.}(2006{\natexlab{a}}){Sana}, {Gosset}, \&
  {Rauw}}]{SGR06_219}
{Sana}, H., {Gosset}, E., \& {Rauw}, G. 2006{\natexlab{a}}, \mnras, submitted

\bibitem[{{Sana} {et~al.}(2006{\natexlab{b}}){Sana}, {Gosset}, {Rauw},
 \& {Vreux}}]{SGR06_elescorial}
{Sana}, H., {Gosset}, E., {Rauw}, G., \& Vreux, J.-M. 2006{\natexlab{b}}, in {X-Ray Universe 2005}, ESA Conf., 6p., in press

\bibitem[{{Sana} {et~al.}(2006{\natexlab{c}}){Sana}, {Naz\'e}, {Gosset},
  {Rauw}, {Sung}, \& {Vreux}}]{SNG06}
{Sana}, H., {Naz\'e}, Y., {Gosset}, E., {et~al.} 2006{\natexlab{c}}, in
  {Massive Stars in Interacting Binaries}, ed. A.~{Moffat} \& N.~{St-Louis},
  ASP Conf. Ser., 5p., in press

\bibitem[{{Schild} {et~al.}(1969){Schild}, {Hiltner}, \& {Sanduleak}}]{SHS69}
{Schild}, R.~E., {Hiltner}, W.~A., \& {Sanduleak}, N. 1969, \apj, 156, 609

\bibitem[{{Schlegel} {et~al.}(1998){Schlegel}, {Finkbeiner}, \&
  {Davis}}]{SFD98}
{Schlegel}, D.~J., {Finkbeiner}, D.~P., \& {Davis}, M. 1998, \apj, 500, 525

\bibitem[{{Seggewiss}(1968{\natexlab{a}})}]{Se68b}
{Seggewiss}, W. 1968{\natexlab{a}}, Zeitschrift fur Astrophysics, 68, 142

\bibitem[{{Seggewiss}(1968{\natexlab{b}})}]{Se68a}
{Seggewiss}, W. 1968{\natexlab{b}}, Veroeffentlichungen des Astronomischen
  Institut der Universitaet Bonn, 79

\bibitem[{{Setia Gunawan} {et~al.}(2002){Setia Gunawan}, {Chapman}, {Stevens},
  {Rauw}, \& {Leitherer}}]{GCS02}
{Setia Gunawan}, D.~Y.~A., {Chapman}, J.~M., {Stevens}, I.~R., {Rauw}, G., \&
  {Leitherer}, C. 2002, private communication

\bibitem[{{Setia Gunawan} {et~al.}(2003){Setia Gunawan}, {Chapman}, {Stevens},
  {Rauw}, \& {Leitherer}}]{GCS03}
{Setia Gunawan}, D.~Y.~A., {Chapman}, J.~M., {Stevens}, I.~R., {Rauw}, G., \&
  {Leitherer}, C. 2003, in IAU Symposium, Vol. 212, A Massive Star Odyssey:
  from main sequence to supernova, ed. K.~{van der Hucht}, A.~{Herrero}, \&
  C.~{Esteban}, 230

\bibitem[{{Shobbrook}(1983)}]{Sh83}
{Shobbrook}, R.~R. 1983, \mnras, 205, 1229

\bibitem[{{Skinner} {et~al.}(2003){Skinner}, {Gagn{\' e}}, \& {Belzer}}]{SGB03}
{Skinner}, S., {Gagn{\' e}}, M., \& {Belzer}, E. 2003, \apj, 598, 375

\bibitem[{{Stickland} \& {Lloyd}(2001)}]{SL01}
{Stickland}, D.~J., \& {Lloyd}, C. 2001, The Observatory, 121, 1

\bibitem[{{Stickland} {et~al.}(1996){Stickland}, {Lloyd}, {Penny}, {Gies}, \&
  {Bagnuolo}}]{SLP96}
{Stickland}, D.~J., {Lloyd}, C., {Penny}, L.~R., {Gies}, D.~R., \& {Bagnuolo},
  W.~G. 1996, The Observatory, 116, 226

\bibitem[{{Stickland} {et~al.}(1997){Stickland}, {Lloyd}, \& {Penny}}]{SLP97}
{Stickland}, D.~J., {Lloyd}, C., \& {Penny}, L.~R. 1997, The Observatory, 117,
  213

\bibitem[{{Str{\" u}der} {et~al.}(2001){Str{\" u}der}, {Briel}, {Dennerl},
  {Hartmann}, {Kendziorra}, {Meidinger}, {Pfeffermann}, {Reppin}, \& {et
  al.}}]{Struder01_pn_mnras}
{Str{\" u}der}, L., {Briel}, U., {Dennerl}, K., {et~al.} 2001, \aap, 365, L18

\bibitem[{{Struve}(1944)}]{Str44}
{Struve}, O. 1944, \apj, 100, 189

\bibitem[{{Sung} {et~al.}(1998){Sung}, {Bessell}, \& {Lee}}]{SBL98}
{Sung}, H., {Bessell}, M.~S., \& {Lee}, S. 1998, \aj, 115, 734

\bibitem[{{Sung} {et~al.}(2004){Sung}, {Bessell}, \& {Chun}}]{SBC04}
{Sung}, H., {Bessell}, M.~S., \& {Chun}, M. 2004, \aj, 128, 1684

\bibitem[{{Turner} {et~al.}(2001){Turner}, {Abbey}, {Arnaud}, {Balasini},
  {Barbera}, {Belsole}, {Bennie}, {Bernard}, \& {et al.}}]{Turner01_mos_mnras}
{Turner}, M.~J.~L., {Abbey}, A., {Arnaud}, M., {et~al.} 2001, \aap, 365, L27

\bibitem[{{van Genderen} {et~al.}(1984){van Genderen}, {Bijleveld}, \& {van
  Groningen}}]{vGBvG84}
{van Genderen}, A.~M., {Bijleveld}, W., \& {van Groningen}, E. 1984, \aaps, 58,
  537

\bibitem[{{Walborn}(1972)}]{Wal72}
{Walborn}, N.~R. 1972, \aj, 77, 312

\bibitem[{{Walraven} \& {Walraven}(1960)}]{WW60}
{Walraven}, T., \& {Walraven}, J.~H. 1960, Bull. Astron. Inst. Netherlands, 15,
  67

\end{thebibliography}

\appendix

\section{On correcting the $L_2$ values in SAS\,v\,5.4.1}\label{app: L2_a}

The equivalent (or transformed) logarithmic likelihood $L_2$  associated with each source candidate detected by the SAS task {\it emldetect} (column DET\_ML in the output file) is given by:
\begin{equation}
L_2=-\ln \left(1- P \left( \frac{ \nu}{2},L' \right) \right)
\label{eq: L2}
\end{equation}
with
\begin{equation}
L'= \sum^{i=n}_{i=1} l_i ,
\label{eq: L'}
\end{equation}

where $P$ is the incomplete Gamma function, $\nu$ is the number of degrees of freedom (d.o.f.) of the fit, $n$ is the number of input images (i.e.\ the number of energy bands times the number of instruments considered), and $l_i = C_i/2$ with $C_i$ being the Cash statistic for image $i$, specially designed by \citet{Ca79} for photon counting experiments. More insight into the physical meaning of Eq.~\ref{eq: L2} will be given in the next section. In this section we focus on the implemented patch for correcting $L_2$ values.

Indeed the logarithmic likelihood $L_2$  is known to be erroneous in SAS version v\,5.4.1 and earlier versions (\xmm\ News \#29 -- 11-Mar-2003). According to SAS Observation Report SASv5.4/8665\footnote{http://xmm.vilspa.esa.es/xmmhelp/}, the factor $2$ in equation $l_i = C_i/2$ has been forgotten, leading to erroneous $L'$ and hence $L_2$. Knowing the number of degrees of freedom $\nu$, it is a simple exercise to invert Eq.~\ref{eq: L2} and to obtain values for $L'$. From Eq.~\ref{eq: L'}, it is obvious that the corrected value for $L'$ is $L'_\mathrm{corr}= L' / 2 $, to be used in  Eq.~\ref{eq: L2} to recover the corrected $L_2^\mathrm{corr}$ value that can then be used for scientific analyses. 

For large values of $L_2$ ($L_2 \ga 10\,000$) the numerical limits of classical compilers are however exceeded. Fortunately Eq.~\ref{eq: L2} tends to a linear relation between $L'$ and $L_2$ for large values and for a given $\nu$. The correction is therefore straightforward with $L_2^\mathrm{corr}=L_2/2$. Though this bug was present at the time we analysed the data, this issue has been fixed later in SAS version \,v\,6.0. We checked our corrected $L_2^\mathrm{corr}$ values against SAS\,v\,6.0 and found them in close agreement.

\section[]{On the choice of coherent detection thresholds using the transformed logarithmic likelihood $L_2$ }\label{app: L2_b}

As it can be deduced from the previous section (App. \ref{app: L2_a}), the logarithmic likelihood $L_2$ is related to the probability that a detected source candidate could be explained by pure random Poissonian fluctuations (and zero count in the source). Computed for each source of the input list, it uses a combination of the Cash statistic $C_i$ obtained for the different input images $i$. The Cash statistic $C_i$ actually results from a likelihood ratio test and obeys a $\chi^2$ distribution \citep{Ca79} with 3 or 4 degrees of freedom (i.e.\ the intensity, the X- and Y-coordinates of the source and, eventually, the extent of the source if allowed). Therefore any linear combination of $n$ $C_i$, and hence any computed $2L'$, also follows a $\chi^2$ statistic with $n+2$ or $n+3$ d.o.f. In this sense, the transformed logarithmic likelihood $L_2$ is indeed linked, through the simple relationship 
\begin{equation}
L_2 = - \ln \left( Q \right), \label{eq: ql2}
\end{equation}
 where 
\begin{equation}
Q=Q\left( \frac{\nu}{2},L'\right)=1- P \left( \frac{ \nu}{2},L' \right), \label{eq: q}
\end{equation}
to the probability $Q$ for a random Poissonian fluctuation to have caused such a high value of $2L'=\sum_{i=1}^{i=n}C_i$ as the one observed. The equivalent logarithmic likelihood $L_2$ will therefore be large if the observed source is likely not produced by a statistical fluctuation, and small otherwise. 

As a consequence, a threshold in $L_2$ can in principle be adopted as a detection limit. However, as we show below, while the expression given in Eq.  \ref{eq: q}  indeed takes into account the number $\nu$ of d.o.f. to compute the $Q$ probability and the subsequent value of $L_2$, it does not allow a direct comparison between $L_2$ obtained with different numbers of input images. This statement is illustrated in the following due consideration.

\begin{table}
\caption{Illustration of consistently determined $L_2$ thresholds (Col. 2) for the different instruments and instrument combinations reported in Col. 1. The number of input images ($n$) and corresponding degrees of freedom ($\nu$) are given in Cols. 3 and 4. $L'$ (Col. 5) is linked to $L_2$ through Eq.~\ref{eq: L2}. A given $L'$ is also linked to other $L'$ of this table through Eq.~\ref{eq: L'} (see text). We emphasize that adopting any of the $L_2$ or $L'$ presented in this table automatically determines the other values of $L'$ and $L_2$ reported here below.}
\label{tab: combL2}.
\centering
\begin{tabular}{c c c c c}
\hline 
\hline
Instr. Comb. & $L_2$ & $n$ & $\nu$ & $L'$ \\
\hline
\mos1 & 10.00 & 3 & 5 & 13.75 \\
\mos2 & 10.00 & 3 & 5 & 13.75 \\
\pn   & 22.77 & 3 & 5 & 27.51 \\
\mos1+\mos2 & 19.25 & 6 & 8 & 27.51 \\
\mos1+\pn   & 31.70 & 6 & 8 & 41.26 \\
\mos2+\pn   & 31.70 & 6 & 8 & 41.26 \\
\mos1+\mos2+\pn & 40.86 & 9 & 11& 55.02 \\
\hline
\end{tabular}
\end{table}

Let us assume that we are dealing, for example, with 3 energy bands and let us only consider point-like source fitting (parameter {\it withextendedsource}='no'). For the purpose of the demonstration, let us adopt a uniform detection threshold, for any instrument or instrument combination, of $L_2=10$. 

As a first step, let us deal with the source detection on the \epicmos1 images. In this particular configuration, there are three input images ($n=3$) that correspond to the three energy bands. From the inversion of Eq.~\ref{eq: L2} with $L_2^{\mathrm{MOS}1}=10$ and $\nu=5$, we obtain $L'_{\mathrm{MOS}1}=13.75$, where $L'_{\mathrm{MOS}1}$ is the sum of the $l^{\mathrm{MOS}1}_i$ for each of the three input images as given by Eq.~\ref{eq: L'}, i.e.\
$$ L'_{\mathrm{MOS}1}=\sum_{i=1}^{i=3} l^{\mathrm{MOS}1}_i .$$
Now assuming that the two instruments \mos1 and \mos2 are exactly identical, a detection threshold $L_2^\mathrm{MOS2}=10$ similarly corresponds to $L'_{\mathrm{MOS}2}=\sum_{i=1}^{i=3} l^{\mathrm{MOS}2}_i=13.75$.

In a next step, let us work with a combination of the two \epicmos\ instruments. Equation \ref{eq: L'} allows us to easily build the combined $L'_{\mathrm{MOS}1+\mathrm{MOS}2}$ as the sum of the $l_i$ for each instrument and energy band :
$$ L'_{\mathrm{MOS}1+\mathrm{MOS}2}=\sum_{i=1}^{i=3} l^{\mathrm{MOS}1}_i + \sum_{i=1}^{i=3} l^{\mathrm{MOS}2}_i = 27.51 .$$

With two instruments and hence 6 images, $ L'_{\mathrm{MOS}1+\mathrm{MOS}2}$ follows a $\chi^2$ distribution with 8 d.o.f. ($\nu=8$). Equation \ref{eq: L2} then gives $L_2^{\mathrm{MOS}1+\mathrm{MOS}2}=19.25$ quite different from the value $L_2^{\mathrm{MOS}1+\mathrm{MOS}2}=10$ obtained with the adopted constant threshold limit $L_2=10$. 

If we consider the use of two identical detectors, the fact, on one side, to combine them and, on the other side, to adopt the same statistical limit for both an isolated detector and a pair of them, allows us to go deeper. Actually, the combined logarithmic likelihood is twice the individual ones:
$$ L'_{\mathrm{MOS}1+\mathrm{MOS}2}=2L'_{\mathrm{MOS1}}=2 L'_{\mathrm{MOS2}}.$$  Thus, this kind of threshold does not preserve the detection limit which is dependent on the particular combination used. If we want to preserve the detection limit adopted for a single instrument, we must, in this example, also multiply the detection threshold by a factor of two, adopting the value 27.51 instead of 13.75 and consequently 19.25 instead of 10 for the transformed $L_2$ statistic.
We can of course extend this result to the \pn\ detector. Making the reasonable assumption that $L'_\mathrm{pn}\approx 2L'_{\mathrm{MOS}}$, a similar reasoning gives $L_2^{\mathrm{pn}}=22.77$, $L_2^{\mathrm{MOS}1+\mathrm{pn}}= L_2^{\mathrm{MOS}2+\mathrm{pn}}=31.70$ and $L_2^{\mathrm{MOS}1+\mathrm{MOS}2+\mathrm{pn}}=40.86$, far from the value of 10.0 initially adopted.  
The intermediate results and numbers of d.o.f. used in establishing these values are given in Table \ref{tab: combL2}.
 Basically, when combining several instruments together, we improve the Poissonian statistics. The fact of adopting a constant value for $L_2$ for various instrumental combinations implies a cut-off in fluxes or count rates that is dependent on the number of detectors considered. Instead, if we prefer to stabilize the cut-off in absolute values of the signal rather independently of the combination used, we have to adapt the $L_2$ value to the situation. \\

In summary, one of the main results of the present discussion is that one should not adopt a constant threshold limit in $L_2$ for different instrument combinations if one wishes to preserve the uniformity of a cut-off level adopted for a given instrument or combination. We have shown that the $L_2$ thresholds in different combinations are linked through Eqs.~\ref{eq: ql2} and \ref{eq: q} and through the detector physical characteristics that condition the $C_i$ values. In consequence, adopting a particular value as a threshold for a specific instrument or instrument combination implicitly assigns related values to the $L_2$ thresholds for any other instrument or combination considered. Therefore, if one wants to adopt a consistent detection threshold whatever the considered instrument or combination are, the previous reasoning becomes a forced step. This issue is particularly relevant to consistently deal with sources that fall on gaps or on specific detector areas where the different instruments do not overlap. We remind that this does not modify the spatial response of the detectors (nor the effect of the field crowdedness). Thus, spatial variations in the effective count rate threshold are still to be expected  and, indeed, they are observed (see Figs.~\ref{fig: pnvsd} and \ref{fig: detlim}).

We finally remind the reader that the above presented method to determine self-consistent $L_2$ thresholds rests on two simplifying, but reasonable, assumptions. The first is that the two \epicmos\ instruments are identical. The second is that the \epicpn\ yields approximately $L'_{\mathrm{pn}}\approx2L'_{\mathrm{MOS}}$. Any refinement of these two assumptions (i.e.\ any relation giving the $L'$ of one instrument as a more realistic function of the $L'$ of the other instruments) can be easily included in the method. This is however beyond the scope of the present discussion. 
The procedure illustrated here has been used in the making of our catalogue. The figures appearing in Table \ref{tab: l2} were indeed established in a similar way (adopting $L_2^\mathrm{MOS1}=L_2^\mathrm{MOS2}=11$) and represent the threshold actually utilized.

\end{document}